\documentclass[twocolumn,tighten,times]{aastex63}

\usepackage{amsmath}
\usepackage[caption=false]{subfig}

\newcommand{\Mtot}{$M_{\mathrm{bin}}$}

\newcommand{\fgd}{$f_{\mathrm{gd}}$}
\newcommand{\vg}{$v_{\rm g}$}

\newcommand{\rhog}{$n_{\rm gd0}$}
\newcommand{\tevol}{$t_{\mathrm{evol}}$}

\shorttitle{Pairing of MBHs in Merger Galaxies}
\shortauthors{Li et al.}

\begin{document}

\title{Pairing of Massive Black Holes in Merger Galaxies Driven by Dynamical Friction}

\author[0000-0002-0867-8946]{Kunyang Li}
\affiliation{School of Physics and Center for Relativistic
  Astrophysics, 837 State St NW, Georgia Institute of Technology,
  Atlanta, GA 30332, USA}   
\email{kli356@gatech.edu}

\author[0000-0002-7835-7814]{Tamara Bogdanovi{\'c}}
\affiliation{School of Physics and Center for Relativistic
  Astrophysics, 837 State St NW, Georgia Institute of Technology,
  Atlanta, GA 30332, USA}
\email{tamarab@gatech.edu}

\author[0000-0001-8128-6976]{David R. Ballantyne}
\affiliation{School of Physics and Center for Relativistic
  Astrophysics, 837 State St NW, Georgia Institute of Technology,
  Atlanta, GA 30332, USA}
\email{david.ballantyne@physics.gatech.edu}




\begin{abstract}
Motivated by observational searches for massive black hole (MBH) pairs at kiloparsec separations we develop a semi-analytic model to describe their orbital evolution under the influence of stellar and gaseous dynamical friction (DF). The goal of this study is to determine how the properties of the merger remnant galaxy and the MBHs affect the likelihood and timescale for formation of a close MBH pair with separation of $\lesssim 1$\,pc.  We compute approximately 40,000 configurations that cover a wide range of host galaxy properties and investigate their impact on the orbital evolution of unequal mass MBH pairs. We find that the percentage for MBH pairing within a Hubble time is larger than 80\% in remnant galaxies with a gas fraction $< 20$\% and in galaxies hosting MBH pairs with total mass $ > 10^6 M_{\rm \odot}$ and mass ratios $\geq1/4$. Among these, the remnant galaxies characterized by the fastest formation of close, gravitationally bound MBHs have one or more of the following properties: (1) large stellar bulge, (2) comparable mass MBHs and (3) a galactic gas disk rotating close to the circular speed. In such galaxies, the MBHs with the shortest inspiral times, which are likely progenitors of coalescing MBHs, are either on circular prograde orbits or on very eccentric retrograde orbits. Our model also indicates that remnant galaxies with opposite properties, that host slowly evolving MBH pairs, are the most likely hosts of dual AGNs at kiloparsec separations.

%
%
\end{abstract}




\section{Introduction}
\label{sec:intro}
Massive black holes (MBHs) with masses $\sim 10^6$--$10^{10} {\rm M}_{\odot}$ are
known to reside in the centers of most massive galaxies \citep{S1982, KR1995,
M1998}. When two galaxies merge, their individual MBHs find themselves orbiting in the gravitational potential of the merger remnant galaxy. Those that form a gravitationally bound MBH binary\footnote{We refer to a system of two MBHs as a {\it MBH pair} before they are gravitationally bound and as a {\it MBH binary} (MBHB) afterwards.}, can become powerful gravitational waves (GWs) sources. The detection of GWs from
stellar-mass black hole binaries by the Laser Interferometer
Gravitational-Wave Observatory \citep[LIGO;][]{LIGO2016} marked the
dawn of the GW era. The first detection of GWs from MBHBs by pulsar timing arrays \citep[PTAs;][]{PTA1990} and the Laser Interferometer Space Antenna
\citep[LISA;][]{LISA2017} is expected to happen in the next $\sim 10-15$ years.
%
%
The expected rate of PTA and \textit{LISA} detections is related
not only to the frequency of galaxy mergers, but also to the physical
processes within the remnant galaxy that bring the individual MBHs to
small enough separations to form a binary. It is therefore important
to understand the evolution of MBHs in post-merger galaxies
in order to anticipate the GW signals probed by the GW observatories.    

 \citet{BBR1980} outlined the framework for calculation of orbital decay of MBH pairs 
following a galactic merger. Depending on the separation and
properties of the merger galaxy, orbital decay of a MBH pair can be
driven by four physical mechanisms. Soon after the merger galaxy
forms, and MBHs are at separations of $\sim 1$ kpc, dynamical friction
(DF) by gas and stars is expected to dominate their orbital decay. The
DF force arises when a massive object, such as a MBH, is moving against a background medium. Gravitational deflection of gas \citep{O1999,KK2007} or collisionless particles (e.g., stars and dark matter) \citep{C1943,AM2012} generates an overdense wake trailing the MBH. The wake exerts gravitational pull onto the MBH, sapping it orbital energy.

The timescale for decay of the MBH orbital separation due to DF is determined by the properties of the two MBHs and their host galaxy. The most important of these include the total mass, mass ratio and initial orbits of the MBHs, and the distribution and kinematics of the gas and stars in the host galaxy. A survey of literature reveals that depending on the exact configuration of a galaxy merger, the timescale for pairing of the MBHs can vary widely, anywhere from $\sim 10^6$ yr to longer than a Hubble time \citep{review2020}. As a result, some merger remnants will promote efficient pairing and subsequently coalescence of MBHs. The others will be unlikely sites of coalescences or may find themselves undergoing multiple galaxy mergers before interactions of triple MBHs can lead to MBH coalescence \citep{Bonetti2018}.

Once the two MBHs are gravitationally bound the stellar "loss-cone" 
scattering is expected to dominate the orbital decay \citep[e.g.,][]{Q1996, QH1997,
  Y2002}. If the galaxy is sufficiently gas reach,  
drag on the binary by the surrounding circumbinary disk may also play an important role for its orbital evolution at separations $\la 0.1$ pc \citep{A2005, D2014, H2014}. When the 
separation falls below $\sim 1000$ Schwarzschild radii GW emission begins to dominate
the orbital decay until coalescence \citep{KT1976, BBR1980}.

A number of earlier studies have explored the pairing of MBHs in
stellar environments using either N-body simulations \citep[e.g.,][]{Q1996,
  QH1997, Y2002, B2006, KJM2011, K2013}, hydrodynamic simulations
of MBHs interacting with gas \citep[e.g.,][]{E2005, D2007,
  C2009}, and semi-analytic models of orbital evolution \citep[e.g.,][]{B2012, Tre2015,K2016, B2016, Tre2018}. These studies
follow the calculation laid out by \citet{C1943}, in which only the
stars moving slower than the MBH contribute to the DF force. An
alternative approach has been presented by \citet{AM2012}, who
calculate the DF force exerted by stars moving both slower and faster
than the MBHs. It is worth noting that none of the earlier
semi-analytic models have examined the effects of DF from gas in the
remnant galaxy. Since mergers of massive galaxies were more frequent
at $z \ga 2$, when gas-rich galaxies were common \citep{Stewart2009, mo2010}, it is pertinent to quantify the impact of gaseous DF in addition to the stars. 

This paper presents a new semi-analytic model of MBH orbital decay due
to DF exerted by both the gas and stars in the remnant galaxy. We
perform a comprehensive exploration of the parameter space using a
suite of approximately 40,000 individual models, defined by the
properties of the MBHs and the remnant galaxy. Based on it, we
evaluate the properties of galaxies that may more efficiently lead to
MBHBs, the progenitors of GW sources, as well as dual AGNs. The rest of this paper is organized as follows. In \S~\ref{sec:methods} we describe the semi-analytic model of the remnant galaxy and the calculation of the DF force. In \S~\ref{sec:results} we present the evolution of eccentricity and inspiral time of the MBH pair as a function of properties of the host galaxy, respectively. We discuss the implications of our findings in \S~\ref{sec:discuss} and conclude in \S~\ref{sec:concl}.

\section{Methods}
\label{sec:methods}

\subsection{Model of a Merger Remnant Galaxy}
\label{sub:galaxymodel}

We assume that a galaxy merger produces a single remnant galaxy containing a stellar bulge, stellar disk,
and a gas disk, all of which contribute the DF force on the MBH
pair. We do not explicitly define the dark matter halo of the galaxy,
as its contribution to the DF force remains $<1\%$ in the central
$1.0$ kpc of the remnant galaxy, for a wide range of merger
configurations considered here. The merger remnant galaxy contains the two MBHs with total mass $M_{\mathrm{bin}}=M_1+M_2$ and mass ratio $q=M_2/M_1$, where $M_1$ and $M_2$ are the masses of the primary and secondary MBH, respectively. The location of the more massive primary MBH (hereafter, pMBH) is fixed at the center of the galaxy throughout the calculation. The orbit of the lower mass secondary MBH (sMBH) is assumed to lie in the plane of the disk of the galaxy remnant. We consider the evolution of a sMBH under the assumption that it was efficiently stripped of all gas and stars that had been bound to it during the galaxy merger (the implications of this assumption is discussed in Sect.~\ref{sec:discuss}). 

The stellar bulge of the galaxy is described by a density profile \citep{BT1987}
\begin{equation}
\label{eq:bulge}
\rho_{b} (r) = \rho_{b_0} \left(\frac{r}{a_b}\right)^{-1.8} e^{-(r/r_b)^2},
\end{equation}
where $\rho_{b_0}$ is the normalization constant chosen so that the
total mass of the stellar bulge is $1000\,M_1$ \citep{M1998}. This
profile is truncated at an outer radius $R_{b} = \log(M_1/10^5\,{\rm M}_{\odot})$~kpc and the two parameters in equation~\ref{eq:bulge} are related to it as $a_b = R_{b}/4$ and $r_b = 1.9R_b/4$. This profile results in the bulge size and properties similar to that of the Milky Way, assuming a corresponding value of $M_1$ \citep[e.g.,][]{BT1987}.

The stellar and gas disks follow the density profile of an exponential disk, which is in the plane of the disk given by \citep{BT1987}:
\begin{equation}
\label{eq:stellardisk}
\rho_{\rm sd} (r) = \Sigma_{\rm sd0}\, e^{-r/R_{\rm sd}} \left(\frac{1}{4z_{\mathrm{thin}}}+\frac{1}{4z_{\mathrm{thick}}}\right)
\end{equation}
and
\begin{equation}
\label{eq:gasdisk}
\rho_{\rm gd} (r) = n_{\rm gd0}\, m_p\, e^{-r/R_{\rm gd}},
\end{equation}
where $\Sigma_{\rm sd0}$ is the surface density of the stellar disk at $r=0$, $n_{\rm gd0}$ is the central number density of the gas disk and $m_p$ is the proton mass. $z_{\mathrm{thin}}$ and $z_{\mathrm{thick}}$ are the scale heights of the thin and thick stellar disks, and $R_{\mathrm{gd}}=2R_{\mathrm{sd}}$ are the characteristic radii of the gas and stellar disks, respectively.  As with the bulge, the characteristic length scales in the stellar disk are assumed to scale with $M_1$, i.e., $R_{\rm sd} = \log(M_1/10^5\ {\rm M}_{\odot})$~kpc, $z_{\mathrm{thin}} = 0.1\,R_{\mathrm{sd}}$, and $z_{\mathrm{thick}} = R_{\rm sd}/3$. To determine $\Sigma_{\rm sd0}$, we first define the gas
disk mass fraction
\begin{equation}
  \label{eq:massfract}
  f_{\rm gd}= M_{\rm gd,1}/(M_{\rm gd,1}+M_{\rm sd,1})\,,
\end{equation}
where $M_{\rm gd,1}$ is the mass of the gas disk within
1~kpc (found by integrating equation~\ref{eq:gasdisk} for a given
$n_{\rm gd0}$). Assuming a value of \fgd\ allows one to calculate
$M_{\mathrm{sd,1}}$, the mass of the stellar disk within 1~kpc, and
then $\Sigma_{\rm sd}$ can be calculated ensuring
equation~\ref{eq:stellardisk} integrates to the correct value.

The temperature profile for the gas disk is calculated using the Toomre's stability criterion \citep{T1964}, which gives the minimum temperature for which the gas disk is stable to gravitational collapse. We set the temperature of the disk to be $10^4$~K above this minimum temperature, since the interstellar medium after the merger of two galaxies is likely to shocked and turbulent \citep[e.g.,][]{BH1991}. Therefore, this gas temperature should be interpret as a proxy for unmodeled turbulence of warm gas. The sound speed in the gas disk is then calculated as $c_{\rm s}= \sqrt{5 k T/3 m_{\rm p}}$. 


\begin{figure}[t!]
\centering 
     \includegraphics[width=0.49\textwidth]{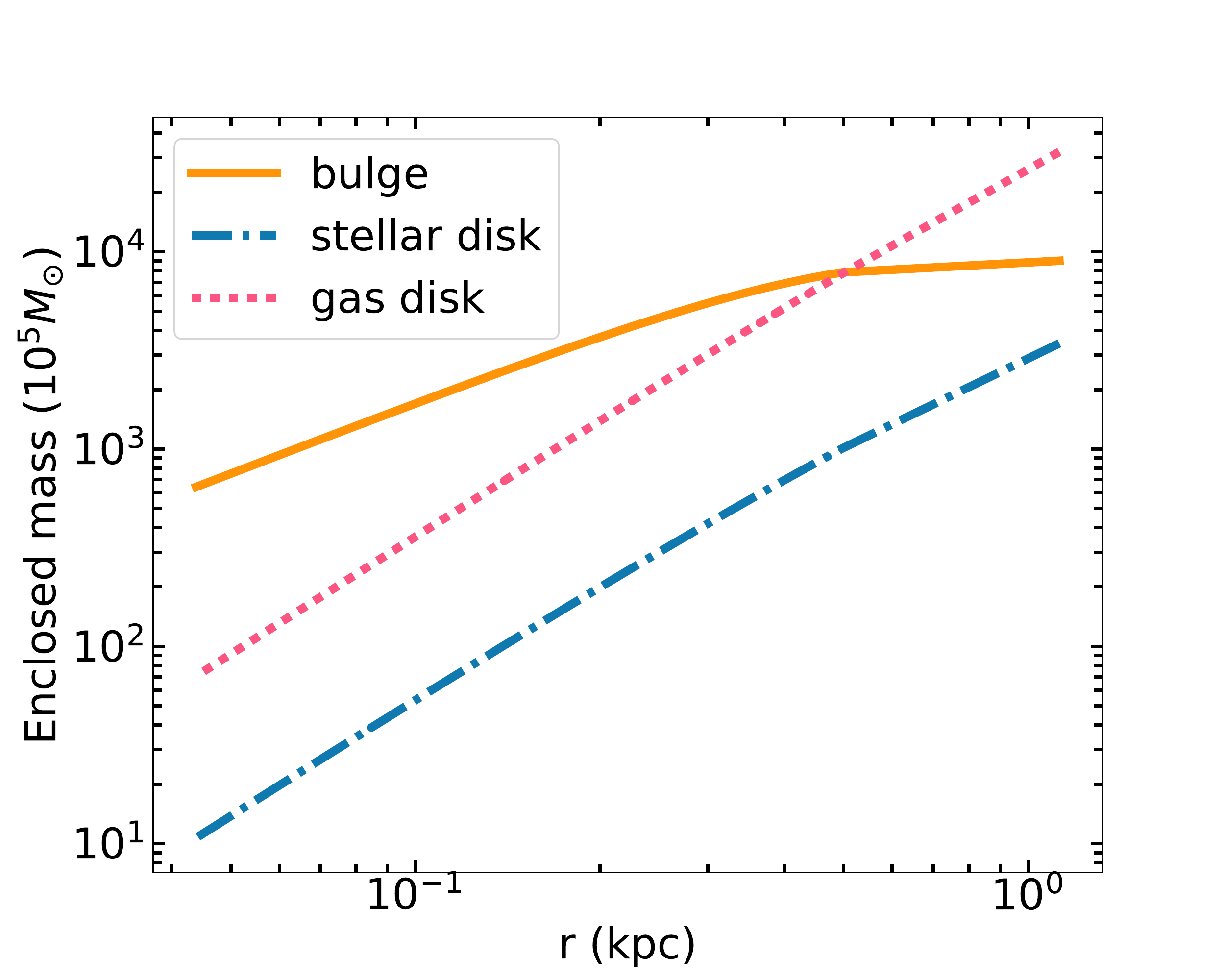}
\caption{Illustration of a mass profile from one of the galaxy models calculated for $M_{\mathrm{bin}}=10^6\ {\rm M}_{\odot}$,  $q=1/9$, $f_{\mathrm{gd}}=0.9$, and $n_{\rm gd0}=300\ {\rm cm}^{-3}$. The enclosed mass comprises contributions from the stellar bulge (solid orange line), stellar disk (blue, dot dashed) and gas disk (pink, dotted).}
\label{fig:massprofile} 
\end{figure}

As described below, the strength of DF on an orbiting MBH depends on the relative velocity between the MBH and galaxy within which it is moving. Therefore, we define the kinematic properties of our galaxy model as follows. We assume the stellar bulge does not have a coherent rotational motion, similar to other semi-analytic models of galaxy mergers \citep{AM2012, KBH2017}. The gas and stellar disks are assumed to be axisymmetric and rotating at the same speed, $v_{\mathrm{g}}(r)$. As the post-merger galaxies may have a range of rotational speeds, DF calculations are performed in galaxy models with $v_{\mathrm{g}}(r)$ varying from
$0.1v_{\mathrm{c}}(r)$ to $0.9v_{\mathrm{c}}(r)$, where
$v_{\mathrm{c}}(r)={\rm \sqrt{r'\frac{d\phi}{dr'}|_{r'=r}}}$ is the circular velocity at radius $r$ for a galaxy with total gravitational potential ${\rm \phi (r)}$. We also consider the impact of
counterrotation on orbital evolution of the sMBH by changing the sign of \vg\ to
negative.

\begin{deluxetable*}{ccC}
\tablenum{1}
\tablecaption{Galaxy Model Parameters\label{tab:params}}
\tablewidth{0pt}
\tablehead{
\colhead{Symbol} & \colhead{Description} & \colhead{Values}
}
\startdata
\Mtot\ & total MBH pair mass & (1,3,5)\times 10^6 \mathrm{M}_{\odot}\\
 	& & (1,3,5)\times 10^7 \mathrm{M}_{\odot}\\
 	 & & (1,3,5)\times 10^8 \mathrm{M}_{\odot}\\
 $q$ & MBH mass ratio & 1/n\,\,  (n=2,\ldots,9) \\
 \rhog\ & central gas number density & 100, 200, 300\,{\rm cm}$^{-3}$ \\
 \fgd\ &  gas disk mass fraction & 0.3, 0.5, 0.9 \\
$v_{\rm g}(r)$ \ & gas disk rotational speed in steps of $0.1 v_{\rm c}(r)$
 & -0.9$v_{\rm c}(r)$,\ldots,0.9$v_{\rm c}(r)$ \\
\enddata
\tablecomments { $v_{\rm g} < 0$ corresponds to the sMBH orbiting in the opposite sense from the gas disk and $v_{\rm g} >0$ indicates a prograde orbit.}
\end{deluxetable*}
%

The mass profile for one of the models is shown in Figure~\ref{fig:massprofile} for illustration. In this example, $M_{\rm bin} = 10^6\,{\rm M}_{\odot}$, $q = 1/9$, $n_{\rm gd0} = 300\ {\rm cm}^{-3}$, and $f_{\rm gd} = 0.9$. The pMBH mass is $M_{1}= 9\times 10^5\ {\rm M}_{\odot}$, and the total mass of the bulge is $9\times 10^8\ {\rm M}_{\odot}$. In this particular galaxy model, the mass of the gas disk dominates over the stellar bulge above a radius of $0.5$~kpc. The stellar disk is less important than both the gas disk and the bulge over all radii due to the large assumed gas fraction of $f_{\rm gd0} = 0.9$.

In summary, our galaxy model is defined by five parameters: \Mtot, $q$, \rhog, \fgd\ and \vg. The parameter grid outlined in Table~\ref{tab:params} corresponds to 39366 model galaxies, for which we calculate the orbital evolution of a sMBH due to DF. The resulting configurations span a wide range in initial orbital eccentricities and include both prograde and retrograde orbits. These models therefore provide a comprehensive view of how DF impacts an orbiting sMBH over a wide range of galaxy types and orbital configurations.

\subsection{Dynamical Friction Force Due to Stars}
\label{sub:sDF}
The evolution of the orbit of sMBH is due to the effects of both stellar
and gaseous DF. The force contributed by the stellar component is exerted
by the bulge and the stellar disk. We compute the DF force due to
stars following the approach laid out in equation~30 in the paper by \citet{AM2012}, which accounts for the DF force contribution from stars moving slower than the sMBH ($v_{\star}<v$) and those moving faster than the sMBH ($v_{\star}>v$, where $v$ is the speed of the sMBH)
\begin{equation}
  \label{eq:bulgeforce}
\vec{F}_{\star} =\vec{F}^{(v_{\star}<v)}_{\star}+\vec{F}^{(v_{\star}>v)}_{\star}.
\end{equation}
If we define $\vec{\mathcal{F}}=-4\pi G^2 M_2^2 \rho_{\star} (r)
(\vec{v}/{v^3})$, where $\rho_{\star}(r)$ is the stellar density given
by either equation~\ref{eq:bulge} or~\ref{eq:stellardisk}, then
\begin{equation}
\label{eq:bulgeforce1}
\vec{F}^{(v_{\star}<v)}_{\star} = \vec{\mathcal{F}} \int_{0}^{v} 4 \pi
f(v_{\star}) v_{\star}^2 \ln \left[\frac{p_{\mathrm{max}}}{G M_2} (v^2
  - v_{\star}^2)\right] dv_{\star}
\end{equation}
and
\begin{eqnarray*}
\vec{F}^{(v_{\star}>v)}_{\star} = \vec{\mathcal{F}} \int_{v}^{v_{\rm esc}} 4 \pi
  f(v_{\star}) v_{\star}^2 \left[\ln \left(\frac{v_{\star} + v}{v_{\star} - v}\right) -2\frac{v}{v_{\star}} \right]dv_{\star}
\end{eqnarray*}
\begin{equation}
\label{eq:bulgeforce2}
\end{equation}
where $f(v_{\star})$ is the velocity distribution of stars, $v_{\rm esc}(r)=\sqrt{-2 \Phi (r)}$ is the escape velocity and $\Phi (r)$ is the gravitational potential of the galaxy remnant. According to equations~\ref{eq:bulgeforce1} and \ref{eq:bulgeforce2}, both the stars moving slower and faster than the sMBH will be deflected into an overdensity wake trailing the MBH, pulling it backward.

We adopt the velocity distribution of bulge stars laid out in equations~11 and 12 in the
paper by \citet{AM2012}
\begin{eqnarray}
  \label{eq:veldisp}
f (v_{\star}) =
\begin{cases}
  0 & v_{\star} > v_{\mathrm{esc}}, \\
  f_0 (2v_c ^2 - v_{\star}^2)^{\gamma - 3/2} & v_{\star} < v_{\mathrm{esc}} 
\end{cases}
\end{eqnarray}
where $\gamma = 1.6$ and the normalization constant $f_0$ is
\begin{equation}
  \label{eq:f0}
  f_0=\frac{\Gamma (\gamma +1)}{\Gamma (\gamma -0.5)}
  \frac{1}{2^{\gamma}\pi^{3/2}v_c^{2\gamma}},
\end{equation}
where $\Gamma$ is the Gamma function. 

For stars in the bulge, we assume that the maximum impact parameter corresponds to the radius of gravitational influence of the pMBH, $p_{\mathrm{max}}= GM_1/\sigma_{\star}^2$. Here $\sigma_{\star}$ is the velocity dispersion of the bulge stars, estimated from the $M-\sigma_{\star}$ relation for the primary MBH \citep{G2009}. This is a conservative assumption that may lead to somewhat longer inspiral times, on average, compared to the cases when $p_{\mathrm{max}}$ is assumed to be position dependent \citep[see][for a recent example]{Bonetti2020}. Comparable MBH mass systems are more affected by this assumption, because in that case $p_{\rm max}$ provides the most restrictive estimate for the DF wake size of the sMBH, whereas configurations with $q < 1/3$ are not significantly affected. This is because most of the DF force is contributed locally and within a few radii of gravitational influence of the sMBH, a region which is well within $p_{\rm max}$ in low $q$ cases \citep{P2017}.


We assume all stars in the stellar disk are rotating with speed \vg, so the velocity distribution is a delta function defined at \vg. The DF from the stellar disk is calculated with Eqs.~\ref{eq:bulgeforce}, \ref{eq:bulgeforce1}, and
\ref{eq:bulgeforce2} where $p_{\mathrm{max}}=GM_1/v_{\rm g}^2$ and $v_{\star}=$\vg.

\subsection{Dynamical Friction Force Due to Gas}
\label{sub:gDF}

We calculate the gaseous DF force exerted onto sMBH following
equations~ 11, 13, 14 in the paper by \citet{KK2007}. This calculation takes into account the contribution to the DF force of the spiral density wake created by the sMBH orbiting in the gas disk of the host galaxy.
\begin{equation}
  \label{eq:gdforce}
\vec{F}_{\rm gd} = -\frac{4\pi (GM_2)^2 \rho_{\rm gd}}{\Delta v^2}(I_R \hat{R}+I_{\phi} \hat{\phi})
\end{equation}
where
\begin{equation} \label{eq:ir}
  I_R =
\begin{cases}
  {\cal M}^2 10^{3.51{\cal M} -4.22} &  {\cal M} < 1.1 \\
  0.5 \ln [9.33{\cal M}^2 ({\cal M}^2 - 0.95)] & 1.1 \leq
  {\cal M} < 4.4\\
  0.3{\cal M}^2 & 4.4 \leq {\cal M}
\end{cases}
\end{equation}
and

\begin{figure*}[t]
\centering
        \begin{tabular}{@{}cc@{}}
            \includegraphics[width=0.49\textwidth]{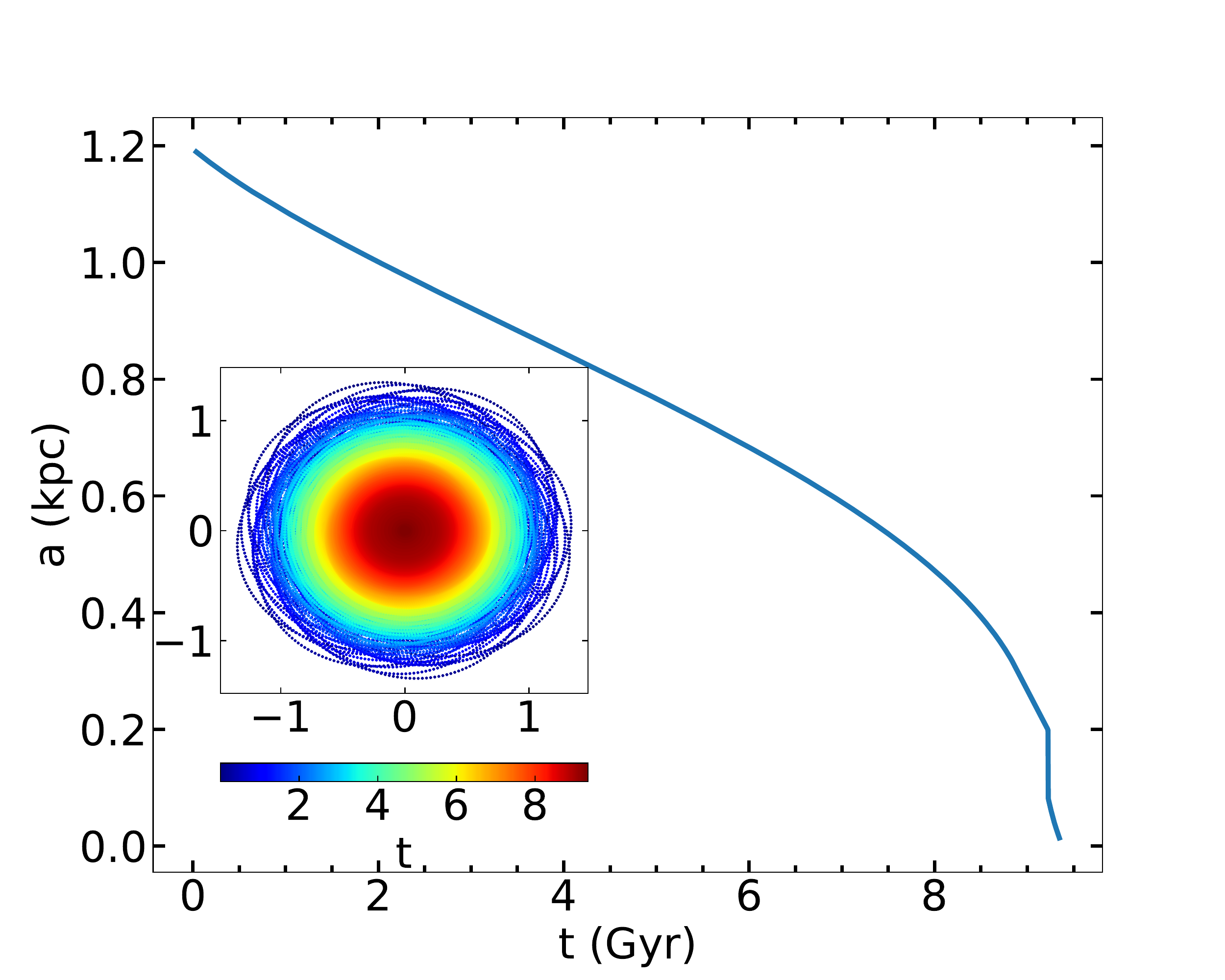}
            \includegraphics[width=0.49\textwidth]{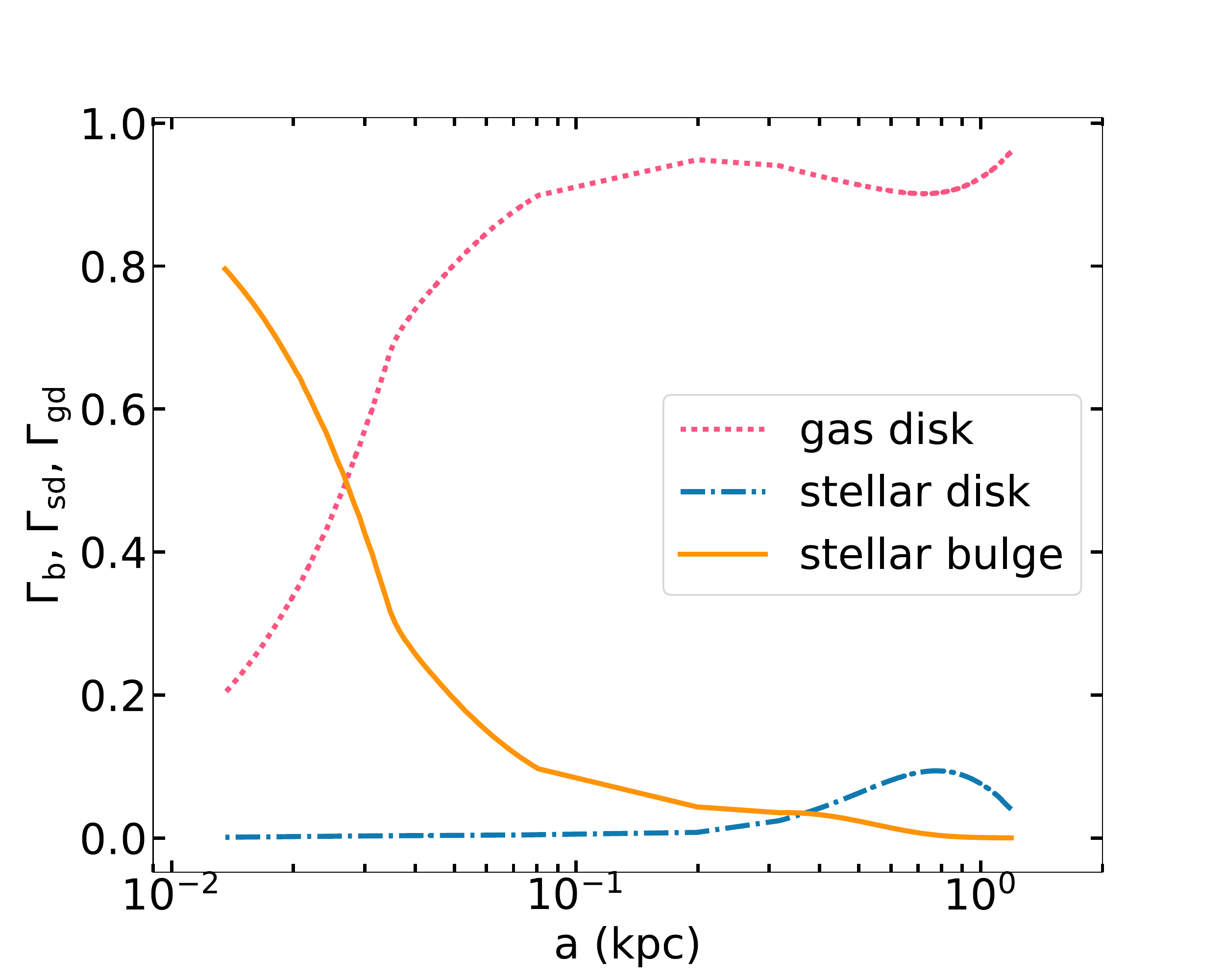}
        \end{tabular}
\caption{{\it Left:} Evolution of the semimajor axis of the MBH pair in a galaxy model calculated for $M_{\rm bin}=10^6\ {\rm M}_{\odot}$, $q=1/9$, $v_{\rm  g}=0.7 v_{\rm c}$, $f_{\rm gd}=0.9$, and $n_{\rm gd0}=300$~cm$^{-3}$. The inset shows the details of orbital evolution of the sMBH. The axis values are in units of kpc and the color bar represents the inspiral time in Gyr. {\it Right:} The fraction of the DF force magnitude contributed by each galaxy component as a function of semimajor axis.}
\label{fig:avstwithinset}
\end{figure*}
%

\begin{equation}
  \label{eq:iphi}
I_{\phi} =
\begin{cases}
0.7706 \ln \left(\frac{1+{\cal M}}{1.0004-0.9185{\cal M}}\right) &
{\cal M} < 1.0 \\
\ln \left [ 330 \left ( \frac{R_p}{r_{\mathrm{min}}} \right ) {\cal
    M}^{-9.58}({\cal M}-0.71)^{5.72} \right ]
& 1.0 \leq {\cal M} < 4.4 \\
\ln \left(\frac{R_p/r_{\mathrm{min}}}{0.11{\cal M}+1.65}\right) & 4.4 \leq {\cal M}
\end{cases}
\end{equation}
where $R_{\rm p}$ is the separation between pMBH and sMBH, and $r_{\rm min}$ is the minimum impact parameter for the gas interacting with the sMBH, which we set to $R_{\rm p}/10$. The ratio $R_p/r_{\mathrm{min}}$ provides a relative measure of the extent of the gaseous wake at any instance of time in the calculation. The wake of maximum extent is bound by the sMBH orbit on one end and the event horizon of the sMBH on the other, but can be smaller than that if the wake is dynamic and its size fluctuates in time. Note that the azimuthal component of the DF force has a weak (logarithmic) dependence on this ratio, so assuming a constant ratio provides a satisfactory approximation.

These expressions above describe the DF force in radial ($R$) and azimuthal ($\phi$) directions, respectively. $I_R$ and $I_{\phi}$ are dimensionless functions of the Mach number, ${\cal M}=
\Delta v/c_{\rm s}$, where $\Delta v$ is the velocity of the sMBH relative to
the gas disk and $c_s$ is the sound speed of the gas.  These expressions imply the radial component of the gaseous DF force that always points towards the center and the azimuthal component that points in the opposite direction from the velocity vector $\Delta \vec{v}$. Both $I_R$ and $I_{\phi}$ peak sharply at ${\cal M} = 1$, so the gaseous DF force is strongest when $\Delta v$ is close to $c_s$. Furthermore, since the strength of $\vec{F}_{\rm gd}$ is proportional to $I_R/\mathcal{M}^2$ and $I_{\phi}/\mathcal{M}^2$ the gaseous DF force is small when the velocity difference between sMBH and the gas disk is large (i.e., $\Delta v \ga 4 c_s$).

\subsection{Calculation of the Orbital Evolution of the sMBH}
\label{sec:orbEvol}

The orbital evolution of the sMBH due to DF is followed until the separation between the two MBHs reaches $1$~pc, at which point the simulation stops. We solve the equation of motion for the sMBH in the $R$ and $\phi$ direction using an $8$th order Runge-Kutta method. The time step is adaptive, and set to less than $1\%$ of the period of a circular orbit, calculated at the instantaneous radius of the sMBH. We determined that the relative error in conservation of energy and angular momentum corresponding to this time step choice is smaller than $0.5\%$, which meets our error tolerance criterion. 

Over the extent of each simulation, the farthest and closest radial
distance between the pMBH and sMBH is recorded for every orbit in
order to estimate the orbital semimajor axis, $a$, and eccentricity,
$e$. As the galaxy potential is not proportional to $1/r$, the orbit
of the sMBH is neither Keplerian nor closed. Thus, the computed $a$
and $e$ are only approximate values used to track the shape and size
of the orbits. Each simulation begins with the sMBH placed at an
initial distance $r_i$ from the pMBH, with an initial velocity of $v_{\rm c} (1+f_{\rm start})^{1/2}$, where $f_{\mathrm{start}}$ is a parameter used to vary the initial eccentricity, $e_{\rm i}$, of the orbit ($e_{\rm i}$ is estimated from the first orbit in the simulation). Throughout the paper, we present the results from  two categories of models: those with low initial eccentricity ($0 \le e_{\rm i} \le 0.2$), initialized with $r_i= 1$~kpc and $f_{\rm start} = 0.5$, and those with high initial eccentricity ($0.8 \le e_{\rm i} \le 0.9$) initialized with $r_i = 0.09$~kpc and $f_{\rm start} = 6$. 

The left panel of Figure~\ref{fig:avstwithinset} illustrates the
evolution of the semimajor axis of a MBH pair calculated using this
approach. In this scenario the separation between the MBHs shrinks
with time. The inset shows the orbit of the sMBH in the galaxy with
the color bar representing the time in the simulation in units of Gyr.

In each simulation, the DF force exerted on the sMBH by the gas disk, stellar disk, and stellar bulge varies with the position of the MBH. We evaluate the relative importance of each of these components by comparing the DF force due to the relevant component averaged over a binary orbit to the orbit-averaged total
DF force. For example, the relative contribution of the gas disk to
the total DF force is calculated as
\begin{equation}
\label{eq:dfactor}
  \Gamma_{\rm gd} = \langle F_{\rm gd} \rangle / \langle F_{\rm tot}
  \rangle.
\end{equation}
Similar expressions are used to compute $\Gamma_{\rm sd}$ and
$\Gamma_{\rm b}$ for the stellar disk and bulge, respectively. The
right panel of Figure~\ref{fig:avstwithinset} shows the relative
contribution of each galaxy component as a function of $a$. For the
orbital decay shown in the left-hand panel, the gas disk contributes to most of the DF force when $a \ga 30$~pc and the stellar bulge dominates at smaller separations. The stellar disk does not contribute significantly to the DF force due to its relatively low mass compared to the bulge and the gas disk (see Figure~\ref{fig:massprofile}).

\section{Evolution of Orbital Eccentricity and Inspiral Time}
\label{sec:results}

\begin{figure*}[t]
\centering
        \begin{tabular}{@{}cc@{}}
            \includegraphics[width=0.49\textwidth]{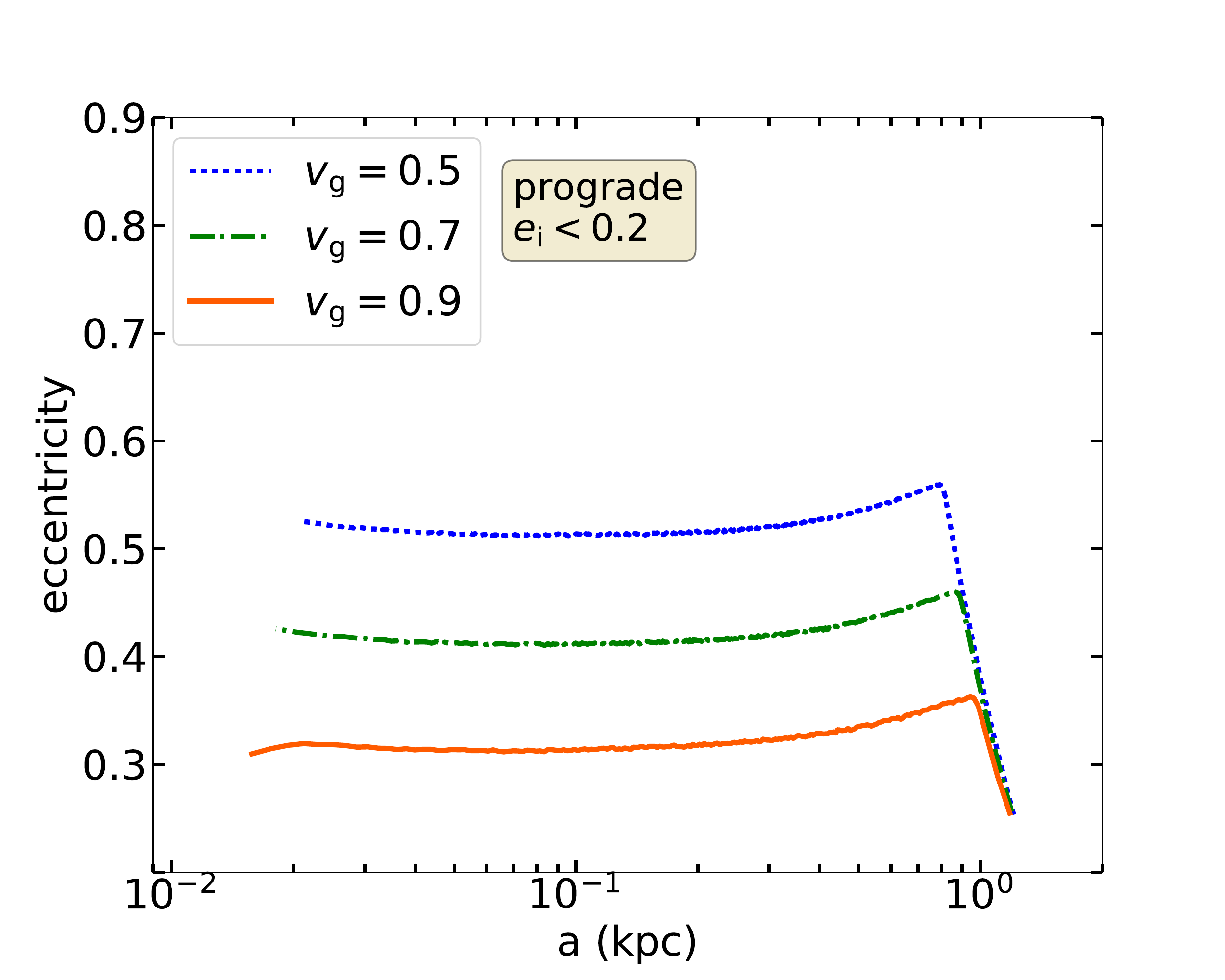}
            \includegraphics[width=0.49\textwidth]{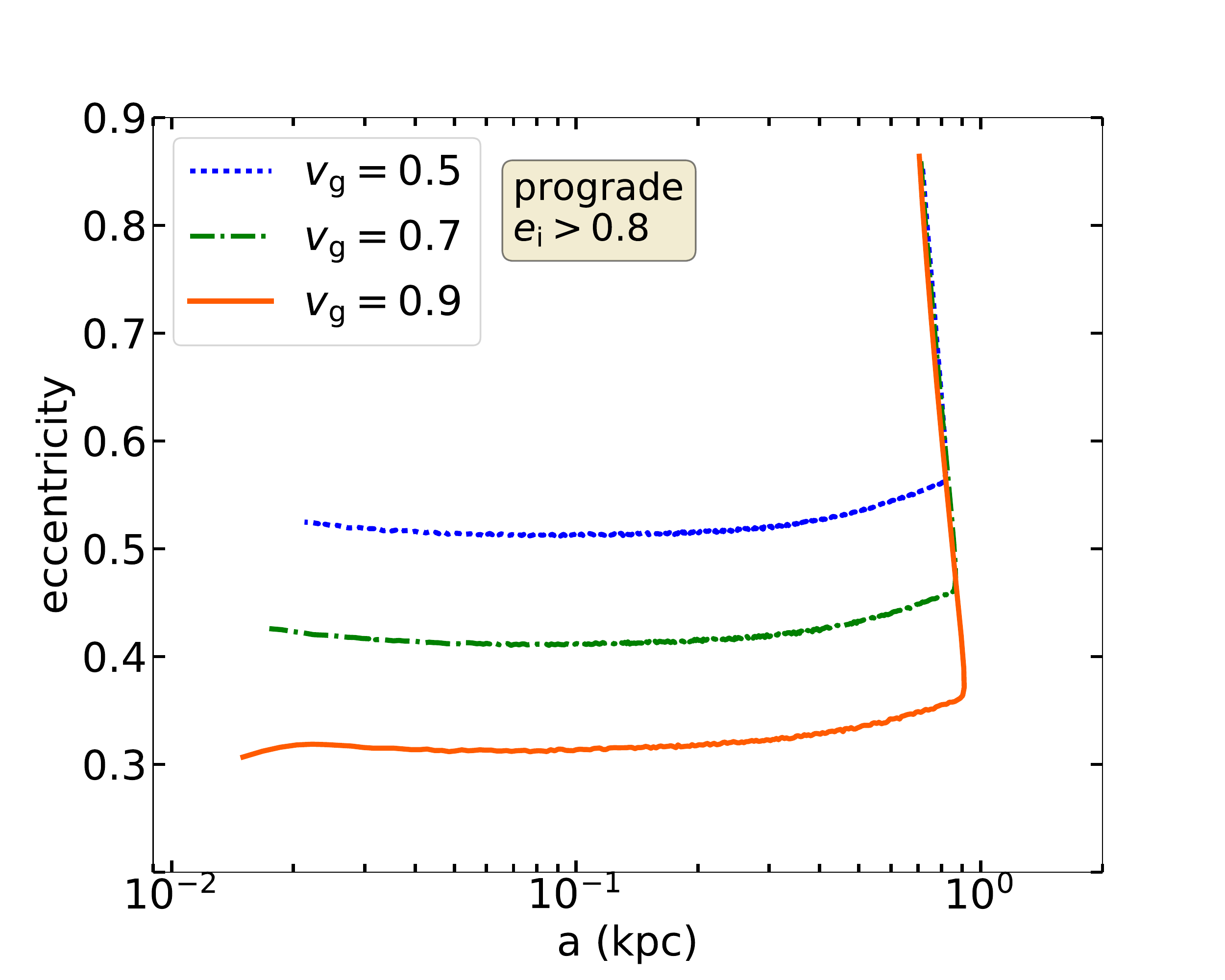}
        \end{tabular}
\caption{Evolution of orbital eccentricity of the sMBH as a function
  of \vg\ in a galaxy where the gas disk is the dominant contributor to the DF force. The models shown are for the sMBH on a prograde orbit and $M_{\rm bin} = 10^7\ {\rm M}_{\odot}$, $q = 1/4$, $f_{\mathrm{gd}}=0.5$, and $n_{\rm gd0}=200$~cm$^{-3}$. The left (right) panel shows configurations with $e_{\rm i} < 0.2$ ($e_{\rm i} > 0.8$). The gas disk speed values in the legends are in units of $v_{\rm c}$.
}
\label{fig:evsa-lowe}
\end{figure*}
%
%

This section describes the evolution of orbital eccentricity and
inspiral time of the sMBH driven by the DF force as a function of the
properties of the remnant galaxy and initial configuration of the MBH
pair. This is of practical interest for two reasons. The first is
that, with all other parameters being the same, more eccentric MBH pairs
that shine as AGNs will be easier to spatially resolve in
observations, because they spend most of the time at large separations,
near the apocenter of their orbit. The second is that MBH pairs that
evolve to become very eccentric have an opportunity to form a
gravitationally bound binary early on. Their small pericenteric
separations lead to interactions with the stars and gas in the dense,
nuclear region of the galaxy and to more efficient gravitational
pairing of MBHs. Such pairs in turn have a greater chance of reaching
the GW emitting regime. In terms of the inspiral time, it is crucial
to understand what types of systems lead to slow inspiral of MBHs, as
these are the ones that will be preferentially detected as dual
AGNs. Similarly, systems conducive to rapid inspiral of the MBHs are
likely to contribute to the population of observed GW sources.

\subsection{Eccentricity Evolution of Prograde and Retrograde Orbits}
\label{S_pro_ret}

In general, the evolution of orbital eccentricity is determined by the action of the DF force on the sMBH integrated along its orbit. Since the DF force is a strong function of the velocity of the sMBH relative to the background medium (see \S~\ref{sub:sDF} and \ref{sub:gDF}), its magnitude and direction can very along an eccentric orbit. Therefore, the DF force can act to either accelerate or decelerate the sMBH at a given point on its orbit, thus driving changes in its orbital eccentricity \citep{Dotti2006}. For example, the eccentricity will increase if
DF causes the sMBH to decelerate at the apocenter of its orbit, but it
will decrease if DF accelerates the sMBH at apocenter. At pericenter, the
effects of DF are reversed and the eccentricity will decrease if the sMBH
is slowed down and will increase if its accelerated. Similarly,
because eccentric sMBHs spend most of the time close to the apocenter,
the overall impact of the DF force will be strongest
there. Consequently, the effect of DF on the eccentricity evolution of
the sMBH depends sensitively on the properties of the remnant galaxy at the radius corresponding to the orbital apocenter of the sMBH \citep{Dotti2006}. 

Figure~\ref{fig:evsa-lowe} shows the impact of different values of
\vg\ on the eccentricity in scenarios when the gas disk (and hence,
its speed and properties) determine the outcome of orbital evolution
of the sMBH. The figure illustrates the evolution for sMBHs on
prograde orbits, split into models with low ($e_{\rm i} < 0.2$) and
high ($e_{\rm i} > 0.8$) initial eccentricity. When $e_{\rm i}$ is low
(left panel of Figure~\ref{fig:evsa-lowe}), the speed of the sMBH at
the apocenter is nearly the circular speed of the galaxy, $v_c$. It
follows that when $v_{\mathrm{g}}=0.5$, 0.7 or $0.9\,v_{\rm c}$, the
relative velocity of the sMBH is along the direction of motion, which
results in a gas DF force dragging the sMBH backwards and slowing it
down. In these scenarios the eccentricity grows until the velocity of
the sMBH becomes approximately equal to the \vg\ at the apocenter and
the Mach number is close to zero. The eccentricity growth stops at
this point, and the evolution enters a phase where there is a relatively slow decay of eccentricity due to gas DF dragging the sMBH backwards at the pericenter. The maximum value of eccentricity reached during the evolution depends on the value of $v_{\rm g}$. When the gas speed is as large as $0.9v_{\rm c}$, the relative speed of the sMBH is small ($\sim 0.1v_{\rm c}$), so the eccentricity growth ends sooner and at a lower eccentricity ($\sim 0.3$). When the gas speed is $0.5v_{\rm c}$, the relative speed of the sMBH is large ($\sim 0.5v_{\rm c}$), and the maximum eccentricity reached is higher ($\sim 0.55$) due to the longer growing phase.

\begin{figure*}[t]
\centering
        \begin{tabular}{@{}cc@{}}
            \includegraphics[width=0.5\textwidth]{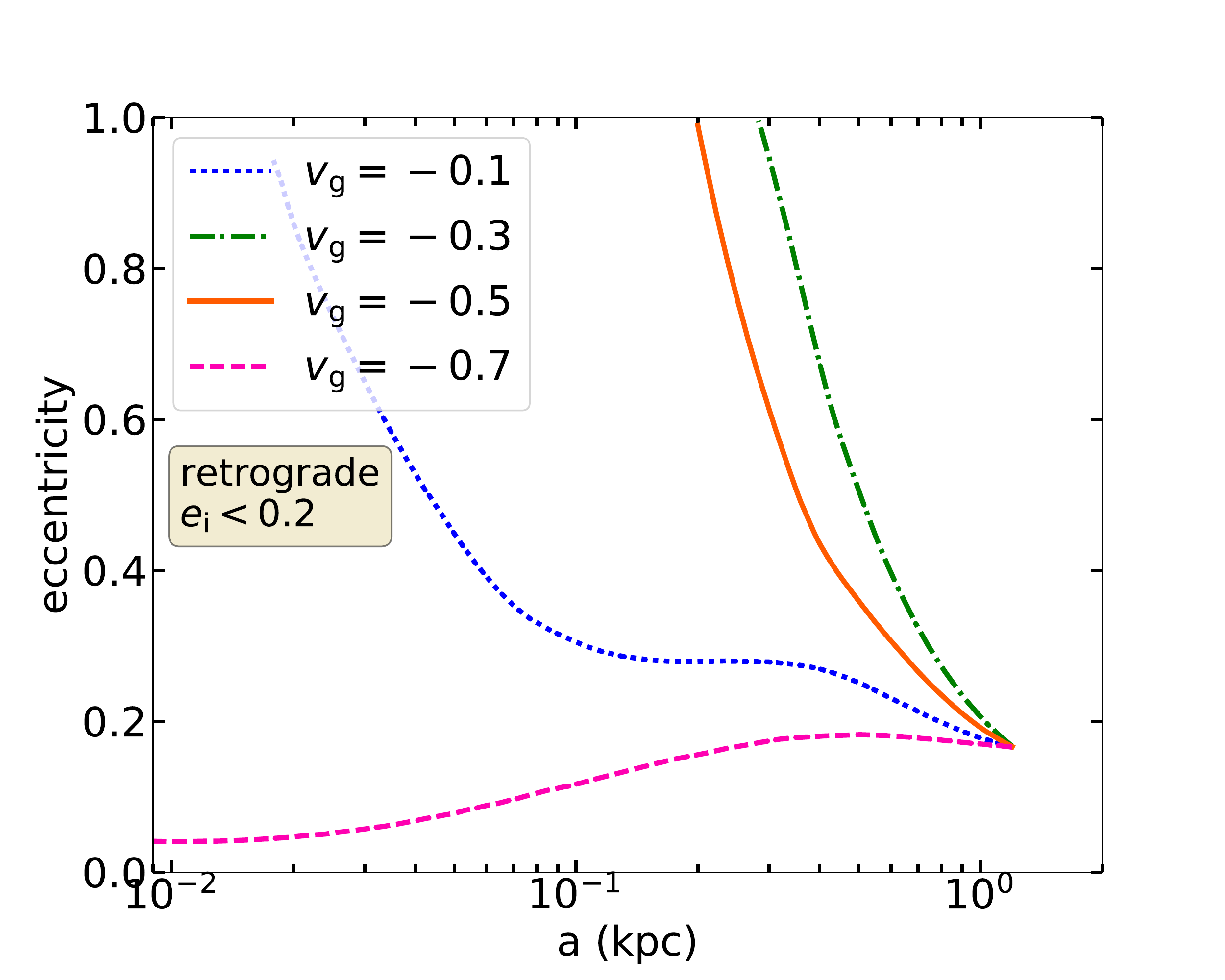}
            \includegraphics[width=0.5\textwidth]{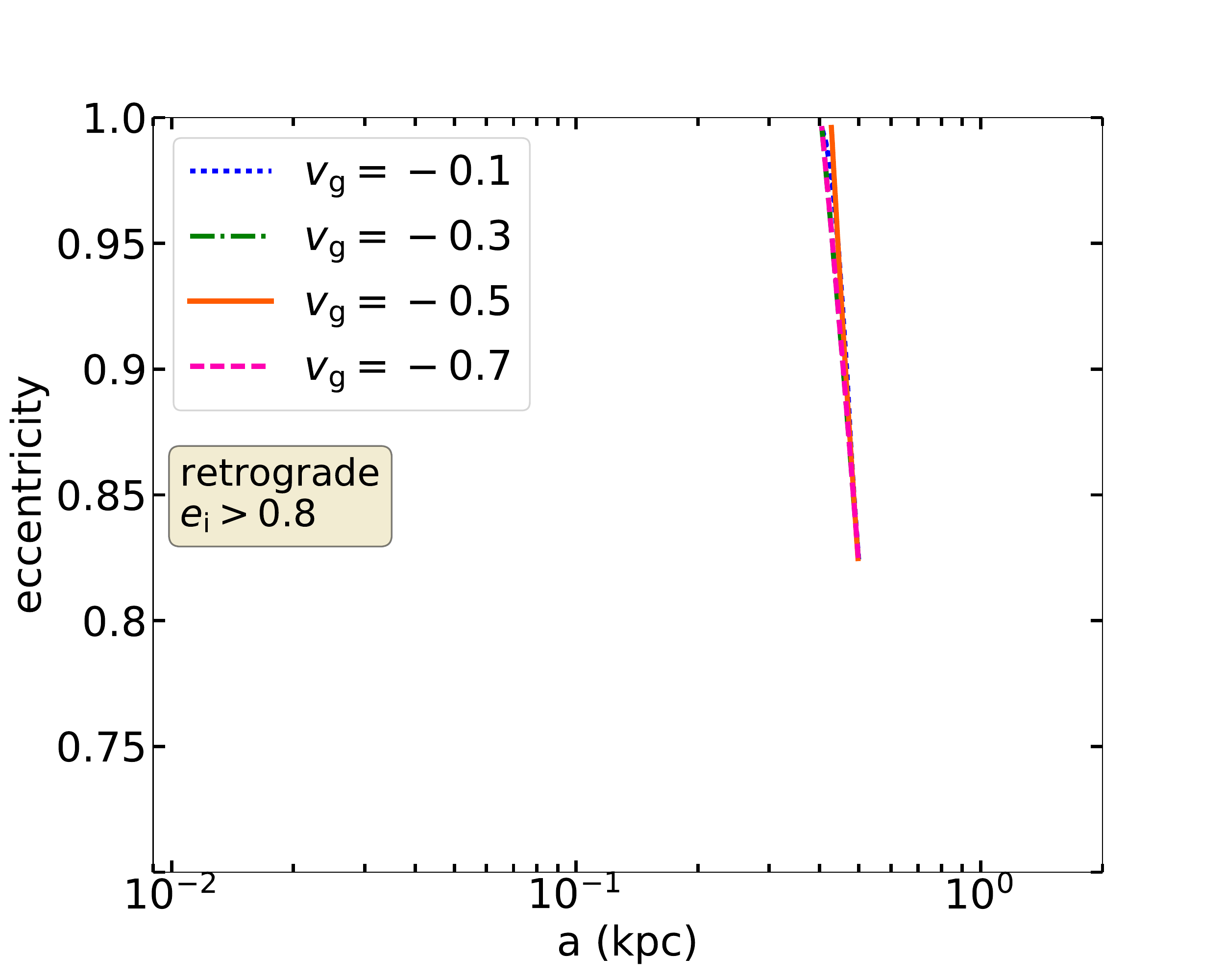}
        \end{tabular}
\caption{Evolution of orbital eccentricity of the sMBH as a function
  of \vg\ in a galaxy where the gas disk is the dominant contributor to the DF force. The models shown are for the sMBH on a retrograde orbit and $M_{\rm bin} = 10^6\ {\rm M}_{\odot}$, $q = 1/4$, $f_{\mathrm{gd}}=0.5$, and $n_{\rm gd0}=200$~cm$^{-3}$. The left (right) panel shows configurations with $e_{\rm i} < 0.2$ ($e_{\rm i} > 0.8$). The gas disk speed values in the legends are in units of $v_{\rm c}$.
}
\label{fig:evsa-lowe_retro}
\end{figure*}

The effect is reversed when $e_{\rm i} > 0.8$ (right panel of
Figure~\ref{fig:evsa-lowe}). In this case, the velocity of the sMBH at apocenter is low relative to the gas disk, and the relative velocity is opposite to the direction of motion of the sMBH.  As a result, the gas DF force now pulls the sMBH forward, speeding it up and decreasing the eccentricity. This continues until the sMBH velocity becomes approximately equal to the \vg\ (the gas DF is close to zero at the apocenter), interrupting the further fast decay of eccentricity. From then on the eccentricity decays relatively slowly due to the gas DF decelerating the MBH at the pericenter. Therefore, we find that for prograde orbits in galactic gas disks, the gas DF always acts to bring the velocity of the sMBH at apocenter close to \vg\, regardless of whether the sMBH initially started on low or high eccentricity orbit. As a consequence, both sets of configurations reach the same final eccentricity, $e_{\rm f}$, with larger values of \vg\ leading to smaller $e_{\rm f}$.

Figure~\ref{fig:evsa-lowe_retro} shows the orbital eccentricity evolution of the sMBHs on retrograde orbits as a function of \vg. As before, we show models with $e_{\rm i} < 0.2$ in the left panel and $e_{\rm i} > 0.8$ in the right panel. For low $e_{\rm i}$ retrograde orbits, when the gas disk rotates faster than $\sim 0.5 v_{\rm c}$ (the dashed pink line), the relative velocity of the sMBH at the apocenter is large which makes the gas DF force inefficient. In the absence of the gas DF, the DF by the stellar bulge takes over the orbital evolution of the sMBH, and as a consequence the orbit circularizes. When the gas disk rotates slower than $\sim 0.5 v_{\rm c}$  however, the gas DF force still dominates the orbital evolution leading to high values of  final eccentricity.

The sMBH on a retrograde orbit with high $e_{\rm i}$ always experiences a backwards gas DF force which slows it down at the apocenter and increases its orbital eccentricity. In these configurations, shown in the right panel of Figure~\ref{fig:evsa-lowe_retro}, the eccentricity continues to grow unchecked until the sMBH plunges below the stopping radius of the simulation.

It is worth noting that this outcome is in part a consequence of the boundary conditions assumed in our calculations. In our model, simulated orbits are bound by $\sim 1$\,kpc on large scales and the plunging radius of 1\,pc, on small scales. As a consequence, the model presented here does not capture a family of retrograde, lowest angular momentum orbits, that correspond to larger fiducial semimajor axes and eccentricities. Such retrograde orbits can evolve to become prograde orbits, if at some point the change in angular momentum per orbit exceeds their orbital angular momentum \citep{B2002, Dotti2006, Mayer2007, C2011, F2013, CD2017, Bonetti2020}.  After the ``orbit flip", these orbits start circularizing as a consequence of the DF torque, which continues to drive growth in their angular momentum. With the exception of orbit flips, we find the same qualitative behavior of eccentric retrograde orbits  (which become more eccentric) and prograde orbits (which circularize), in agreement with these works. This gives us confidence that our model reproduces the salient features of dynamical evolution of the sMBH within the parameter range of interest of this calculation.

\subsection{Evolution of Eccentricity in Gas Disk vs Bulge Dominated Scenarios}

\label{subsub:finalecc}
\begin{figure*}[t]
\centering
        \begin{tabular}{@{}cc@{}}
            \includegraphics[width=0.5\textwidth]{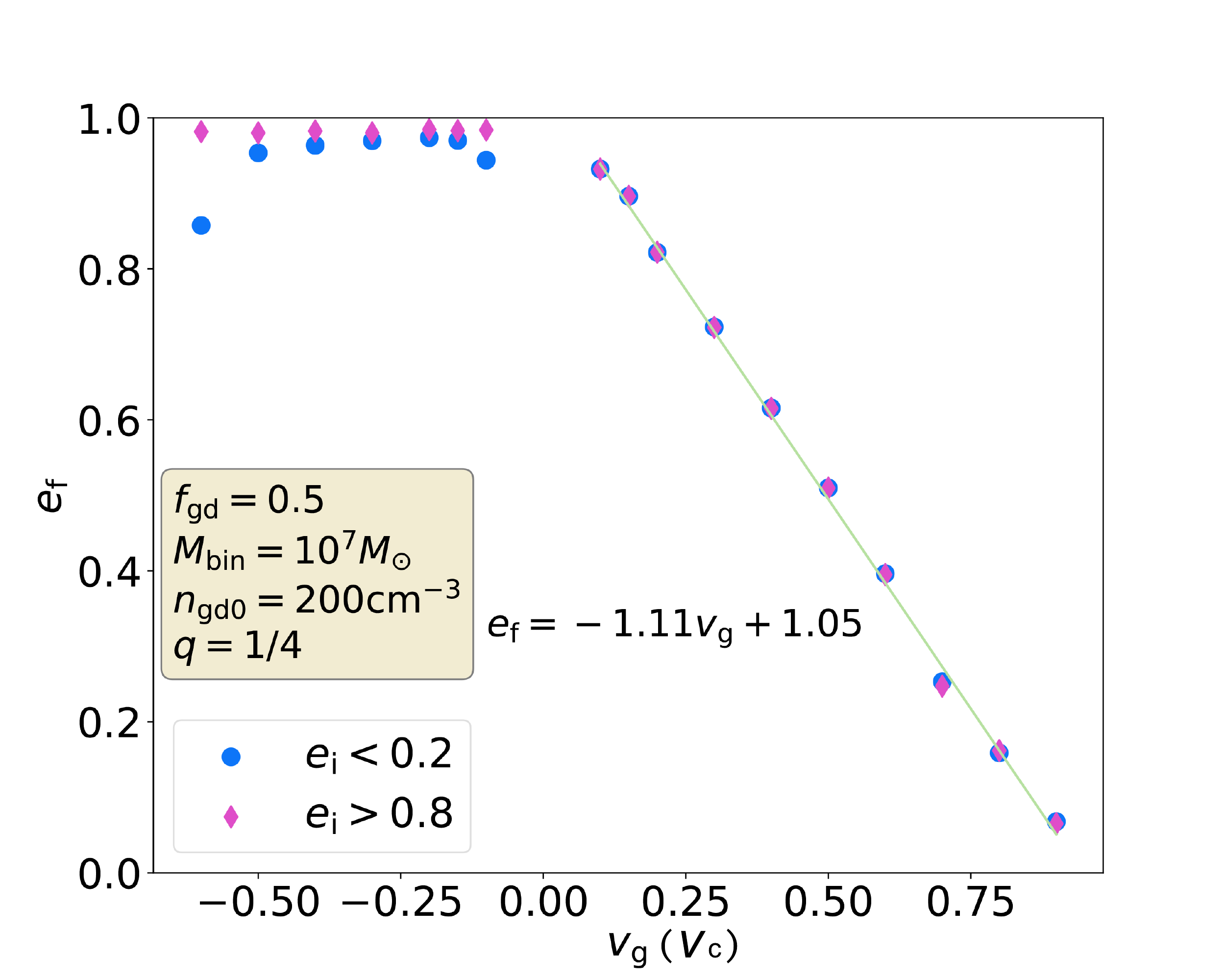}
            \includegraphics[width=0.5\textwidth]{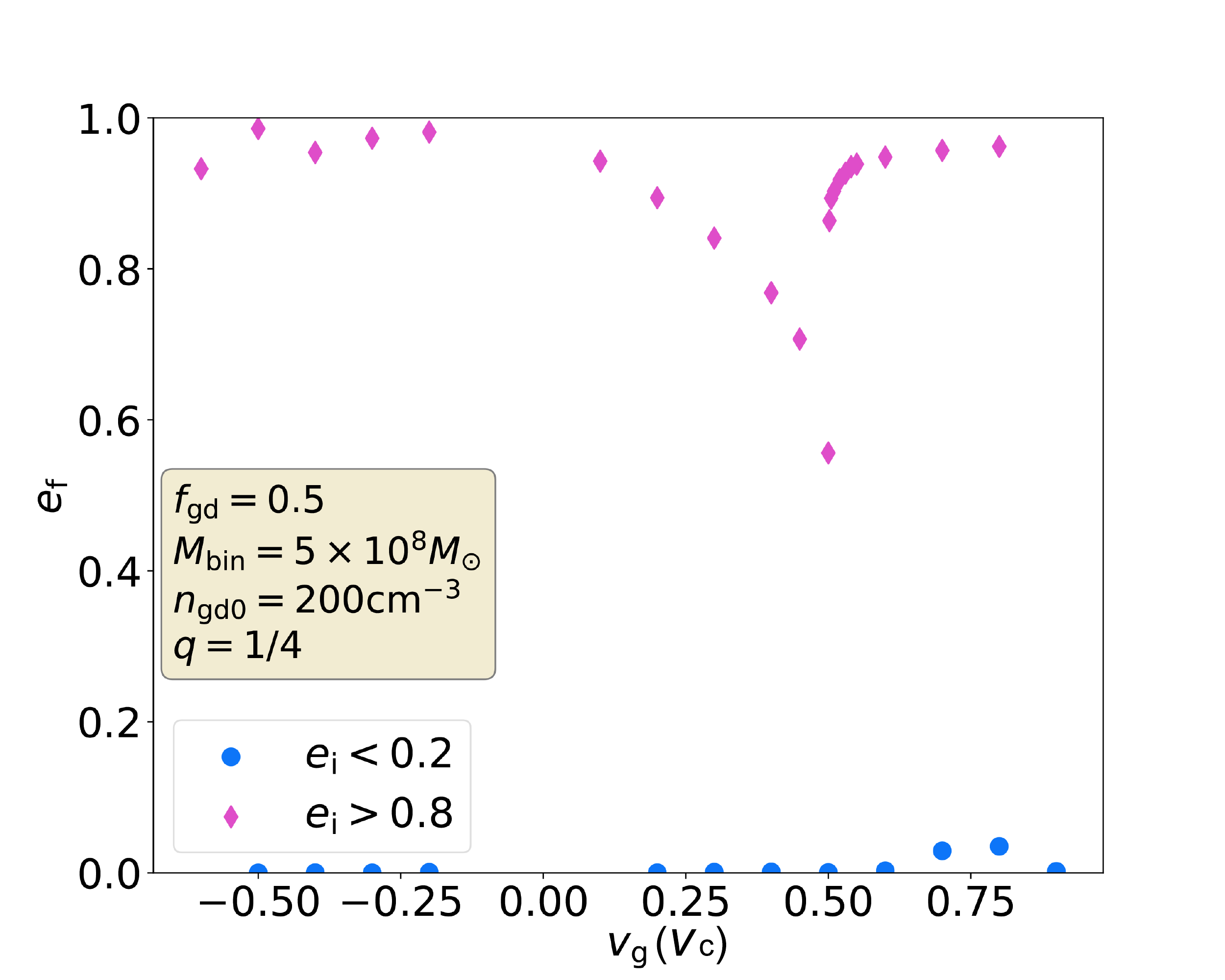}
        \end{tabular}
\caption{{\it Left:} The final orbital eccentricity of the sMBH, $e_{\rm f}$ in scenarios where the gas disk is the main contributor to the DF force and $M_{\mathrm{bin}} = 10^7\ M_{\odot}$, $q = 1/4$, $n_{\rm gd0} = 200$~cm$^{-3}$, $f_{\mathrm{gd}} = 0.5$. The solid green line is the best fit for models with prograde orbits ($v_g > 0$), whose functional form is indicated in the plot. {\it Right:}  The final orbital eccentricity of the sMBH, $e_{\rm f}$ in scenarios where the stellar bulge is the main contributor to the DF force. The model parameters are identical to those in the left panel with the exception of $M_{\rm bin} = 5 \times  10^8\ M_{\odot}$. In both panels, blue circles (pink diamonds) represent orbits with $e_{\rm i} < 0.2$ ($e_{\rm i} > 0.8$).}
\label{fig:linea}
\end{figure*}

\begin{figure*}
  \includegraphics[width=0.49\textwidth]{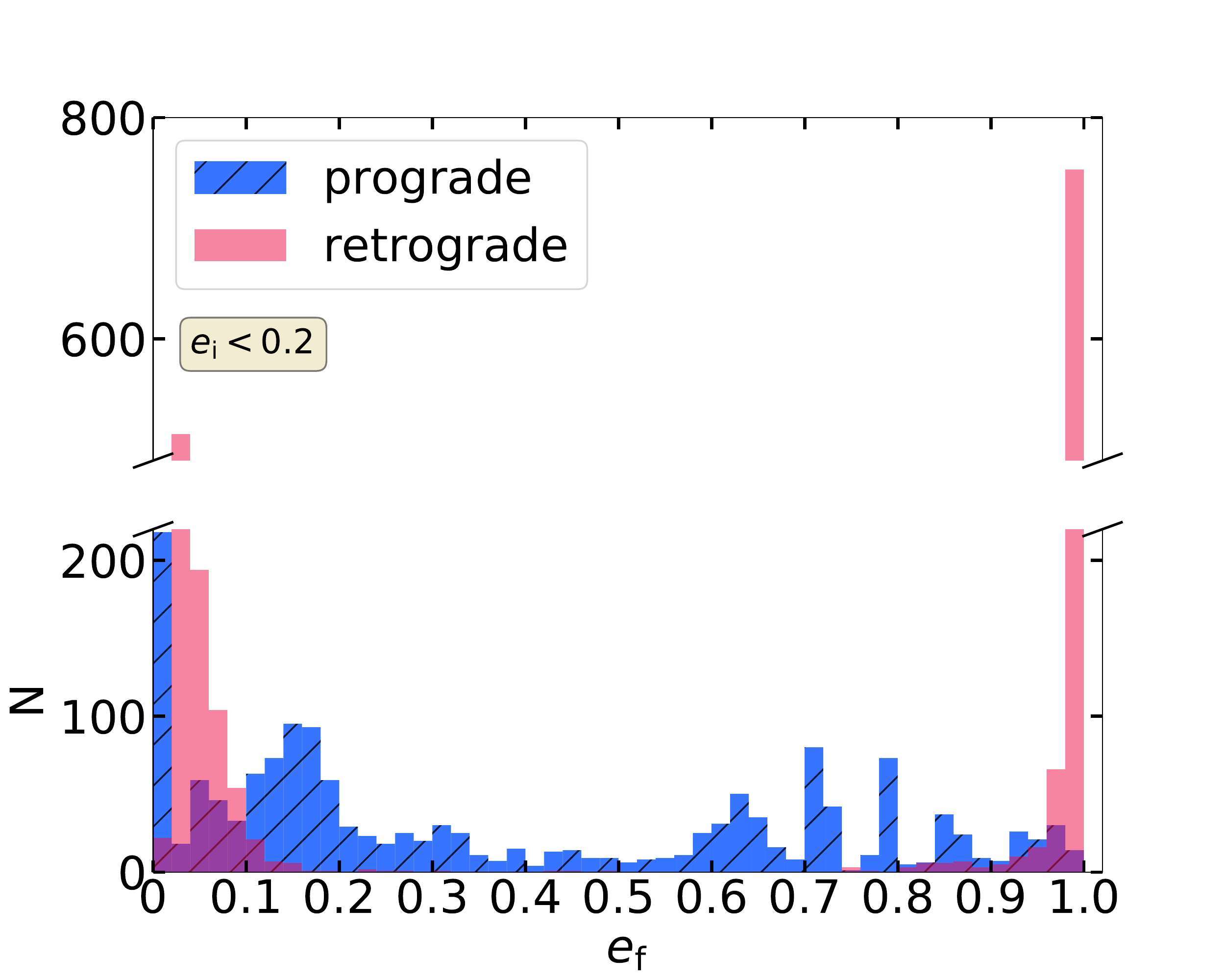}
  \includegraphics[width=0.49\textwidth]{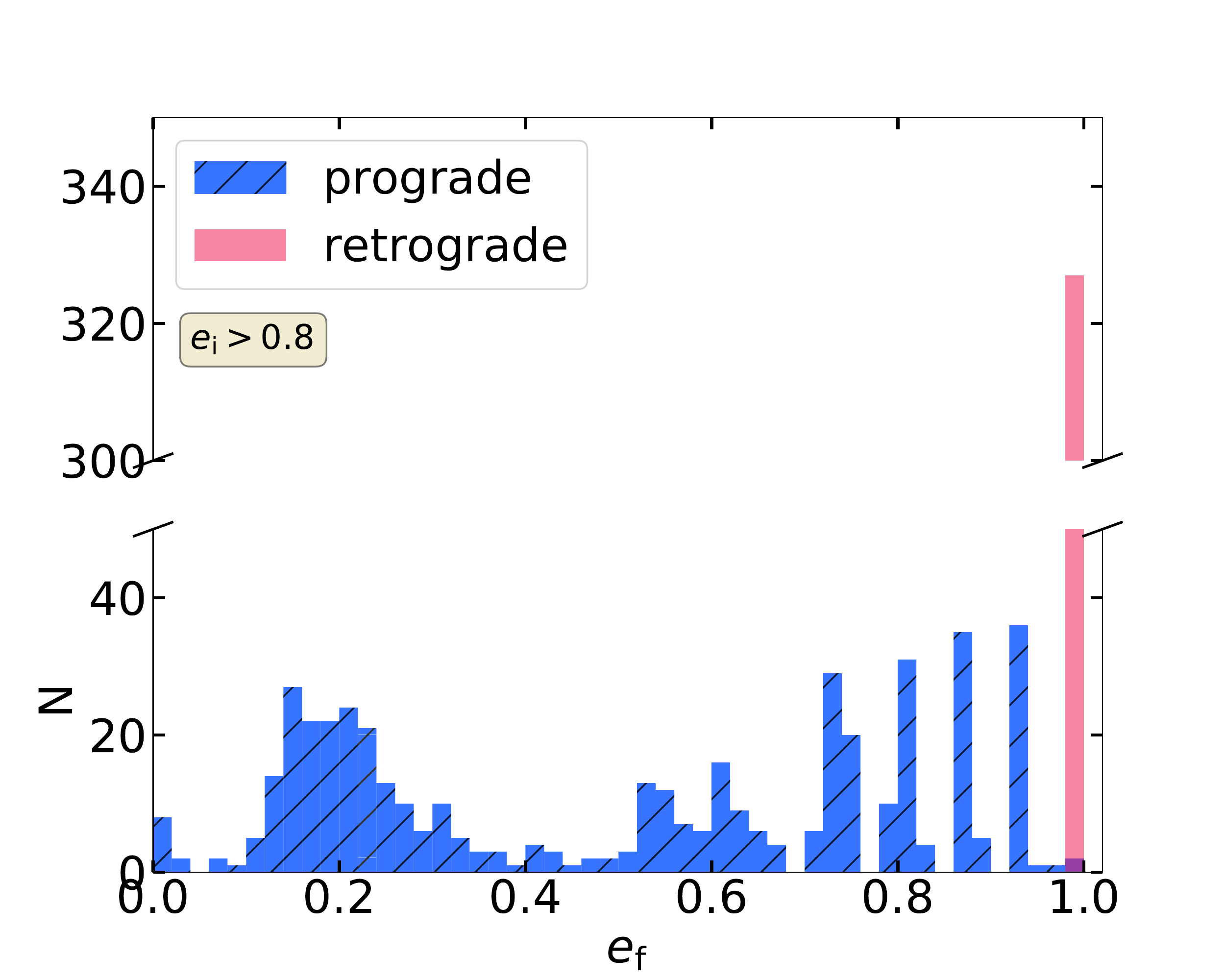}
  \caption{Histograms of $e_{\rm f}$ computed from the entire model
    suite for prograde
    and retrograde orbits with low (left) and high $e_{i}$ (right).}
  \label{fig:efhisto}
\end{figure*}

%
%

The left panel of Figure~\ref{fig:linea} shows the values of $e_{\rm f}$ for the prograde and retrograde orbits in a galaxy where the gas disk is the dominant contributor to the DF force. For prograde cases ($v_g > 0$),
$e_{\rm f}$ depends linearly on \vg\, regardless of the initial eccentricity. In this regime, the two are characterized by a relation $e_{\rm f} = (-1.11\pm 0.02)v_{\rm g} + (1.05\pm 0.01)$. For retrograde orbits ($v_g < 0$), $e_{\rm f}$ is always large and close to 1, since the gas disk is rotating in the opposite sense to the sMBH, driving the eccentricity growth. For orbits with $e_{\rm i} < 0.2$, the gaseous DF force starts to
weaken when $v_{\rm g} \lesssim -0.6 v_{\rm c}$, and so $e_{\rm f}$ does not reach $1$. In this regime, the relative velocity between sMBH and the gas disk is sufficiently large that the gas DF force becomes inefficient at the apocenter, as described in \S~\ref{S_pro_ret}.

The eccentricity evolution proceeds entirely differently in galaxies where orbital evolution of the sMBH is dominated by the stellar bulge, as opposed to the gas disk. The right panel of Figure~\ref{fig:linea} shows relevant models with a more massive, dominant stellar bulge with $M_{\rm bin} = 5\times 10^8\ M_{\odot}$  and sMBH orbits with initially low eccentricity, $e_{\rm i} < 0.2$. In all models with these properties, the orbit of sMBH is circularized for any value of \vg, including both prograde and retrograde orbits.  This outcome is expected for bulges described by stellar density profiles with power law indices $\alpha_b > 1.5$ \citep{Q1996,GQ2003,AM2012}. Since in this study we choose $\alpha_b = 1.8$, the DF force due to the bulge leads to low $e_{\rm f}$.  

In contrast to the low eccentricity orbits, the stellar bulge does not
have a strong impact on orbits with $e_{\rm i} > 0.8$ (marked by pink
diamonds in the right panel of Figure~\ref{fig:linea}). This is
because the sMBH on a highly eccentric orbit spends most of its time
close to the apocenter, where contribution of the bulge to the DF
force is negligible. For example, the largest stellar bulge in our model suite
(corresponding to $M_{\rm 1}=9\times10^7 M_{\rm \odot}$) affects
the evolution of the sMBH orbit to at most $\sim 300$~pc. This
leaves the gas disk as the main contributor to the DF force at larger distances for almost
all high eccentricity systems. Since the gas disk is under-massive relative to the bulge
in the models shown in the right panel of  Figure~\ref{fig:linea}, its contribution to the DF force is relatively small. This leads to a weak eccentricity evolution and high final eccentricities for almost all values of \vg, except at $v_{\rm g} \approx 0.5 v_{\rm c}$, where the gas DF friction force peaks because ${\cal M} \approx 1$.

Figure~\ref{fig:efhisto} shows histograms of the final eccentricity
from out entire model suite for both prograde and retrograde orbits
with $e_{\rm i} < 0.2$ (left) and $e_{\rm i} > 0.8$ (right
panel). Both panels show a similar distributions of orbits with
$e_{\rm f} > 0.1$, whose evolution is always dominated by DF in the
gas disk. The difference is that a substantial fraction of models in
which the sMBHs initially has low eccentricity tend to further
circularize, as represented by the histogram peak at $e_{\rm f} <
0.1$, whereas in the right panel of Figure~\ref{fig:efhisto} these
models are absent. This distinction arises because in models with low
$e_{\rm i}$, the evolution of orbits can be dominated by the stellar
bulge over a substantial portion of time, leading to efficient
circularization. Evolution of the high $e_{\rm i}$ orbits is dominated
by the DF in the gas disk, which does not often lead to
circularization.

More specifically, for models with low $e_{\rm i}$, the circularization fraction of prograde and retrograde orbits is 23.4\% and 49.0\%, respectively\footnote{We refer to orbits with $e_{\rm f} < 0.1$ as {\it circularized}.}. The retrograde orbits are twice as likely to be circularized because they experience even less gas DF relative to the prograde orbits. As a consequence, the evolution of retrograde orbits is likely to be bulge dominated and to lead to circularization. In comparison, for models with high $e_{\rm i}$, the circularization fraction of prograde and retrograde orbits is 2.8 $\%$ and 0.0 $\%$, respectively.  

These results indicate that the sMBHs on initially highly eccentric orbits tend to remain highly eccentric if they are (a) on retrograde orbits or (b) orbiting in gas poor galaxies, so that the gas has no chance to strongly affect the evolution of eccentricity. The implication of this result is that a sMBH on a highly eccentric orbit will spend most of its time at larger distances from the pMBH (close to the orbital apocenter), making a detection of this pair as a dual AGN more likely, if they are both active. We therefore anticipate that catalogs of observed dual AGNs will be biased towards systems that satisfy one of the above criteria.

\begin{figure*}[t]
\centering
        \begin{tabular}{@{}cc@{}}
            \includegraphics[width=0.49\textwidth]{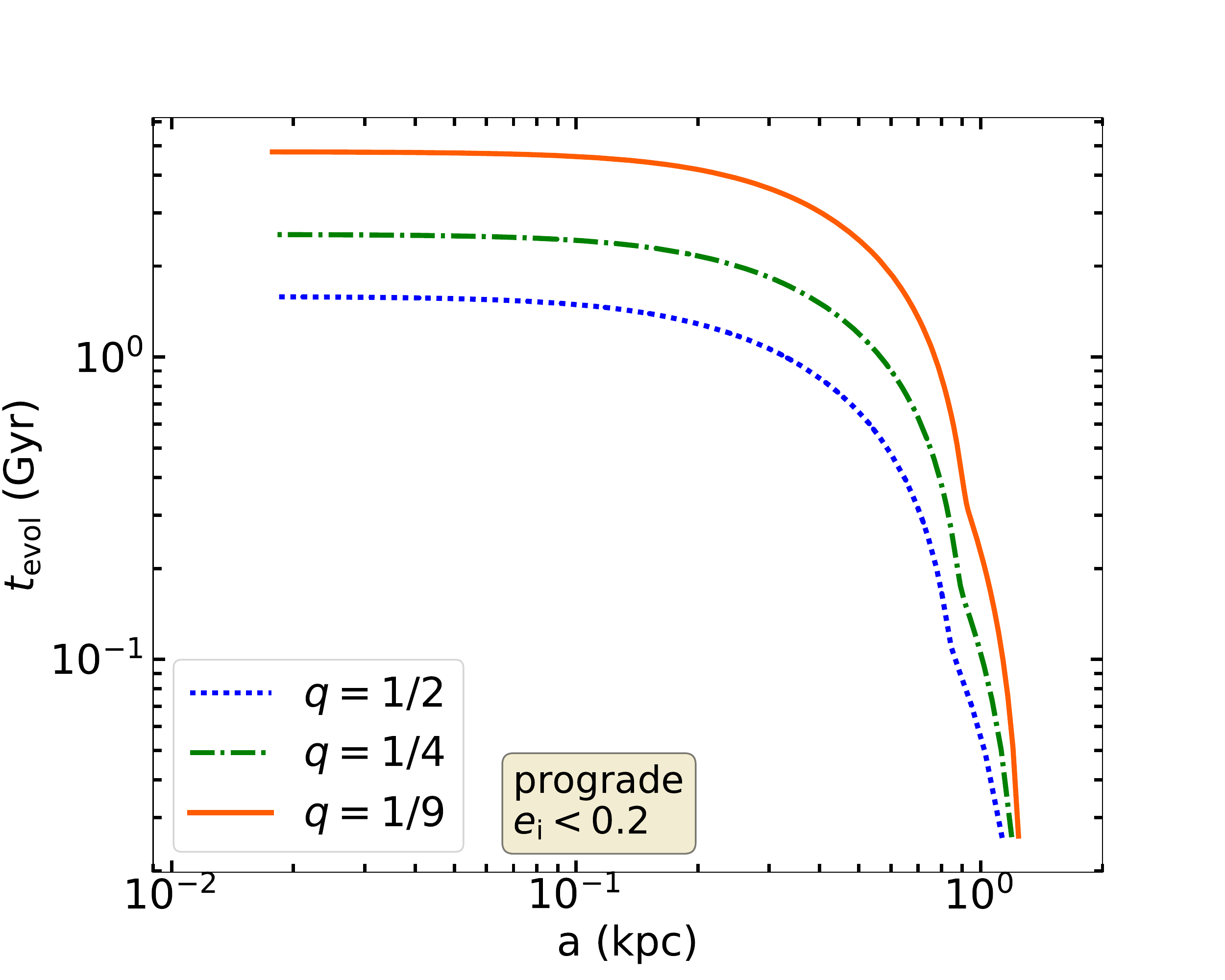}
            \includegraphics[width=0.49\textwidth]{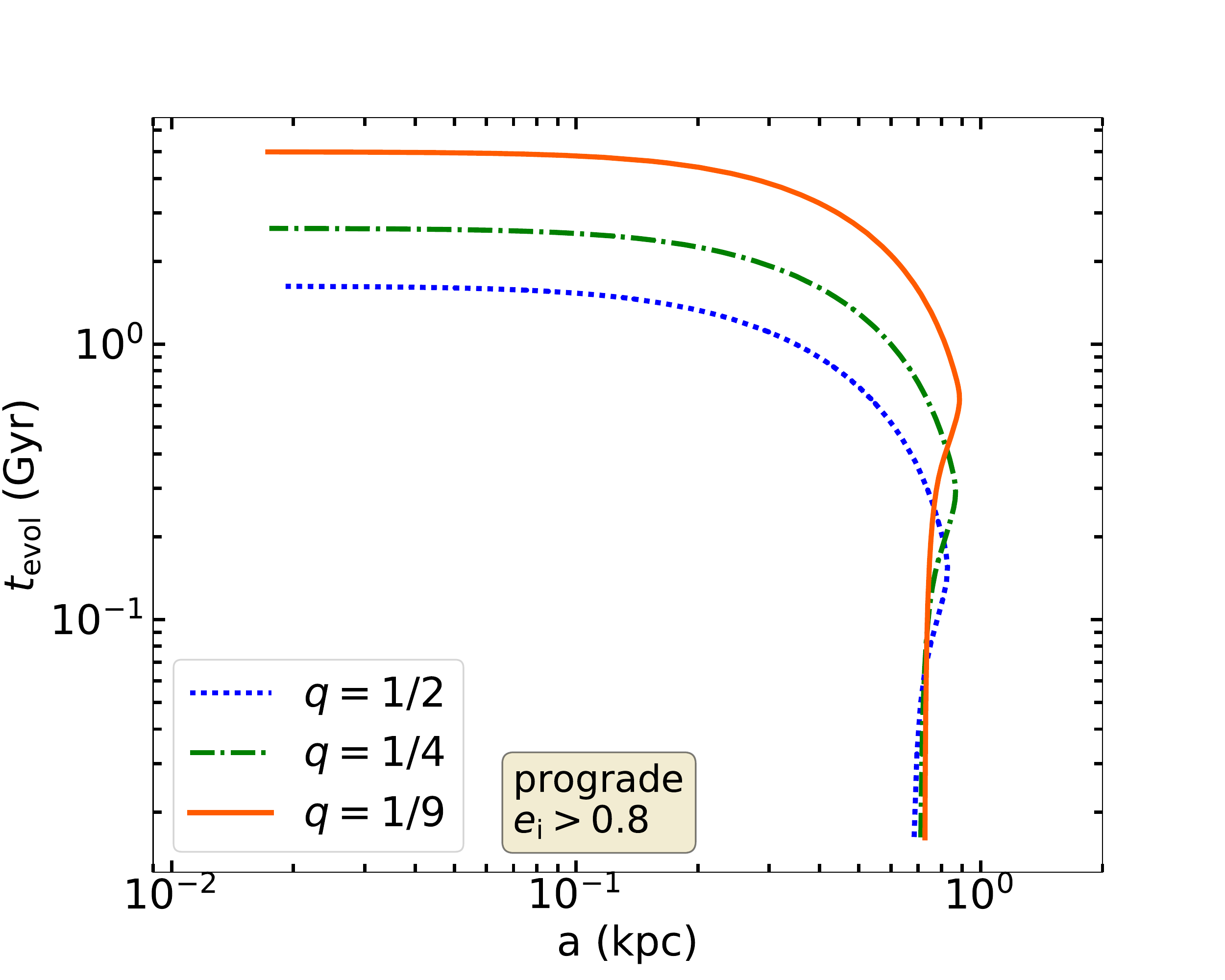}
        \end{tabular}
\caption{Evolution of the semimajor axis for sMBHs with low (left) and high $e_{\rm i}$ (right) in a remnant galaxy with $M_{\mathrm{bin}}=10^7\ {\rm M}_{\odot}$,  $n_{\rm gd0}=200$~cm$^{-3}$, $f_{\mathrm{gd}}=0.5$ and $v_g=0.6v_c$. Different lines mark the mass ratio of the MBH pair, $q = 1/9$ (solid red), $q = 1/4$ (dot-dashed green) and $q = 1/2$ (dotted blue).}
\label{fig:tvsa-lowe}
\end{figure*}

\begin{figure*}[t]
\centering
     \includegraphics[width=\textwidth]{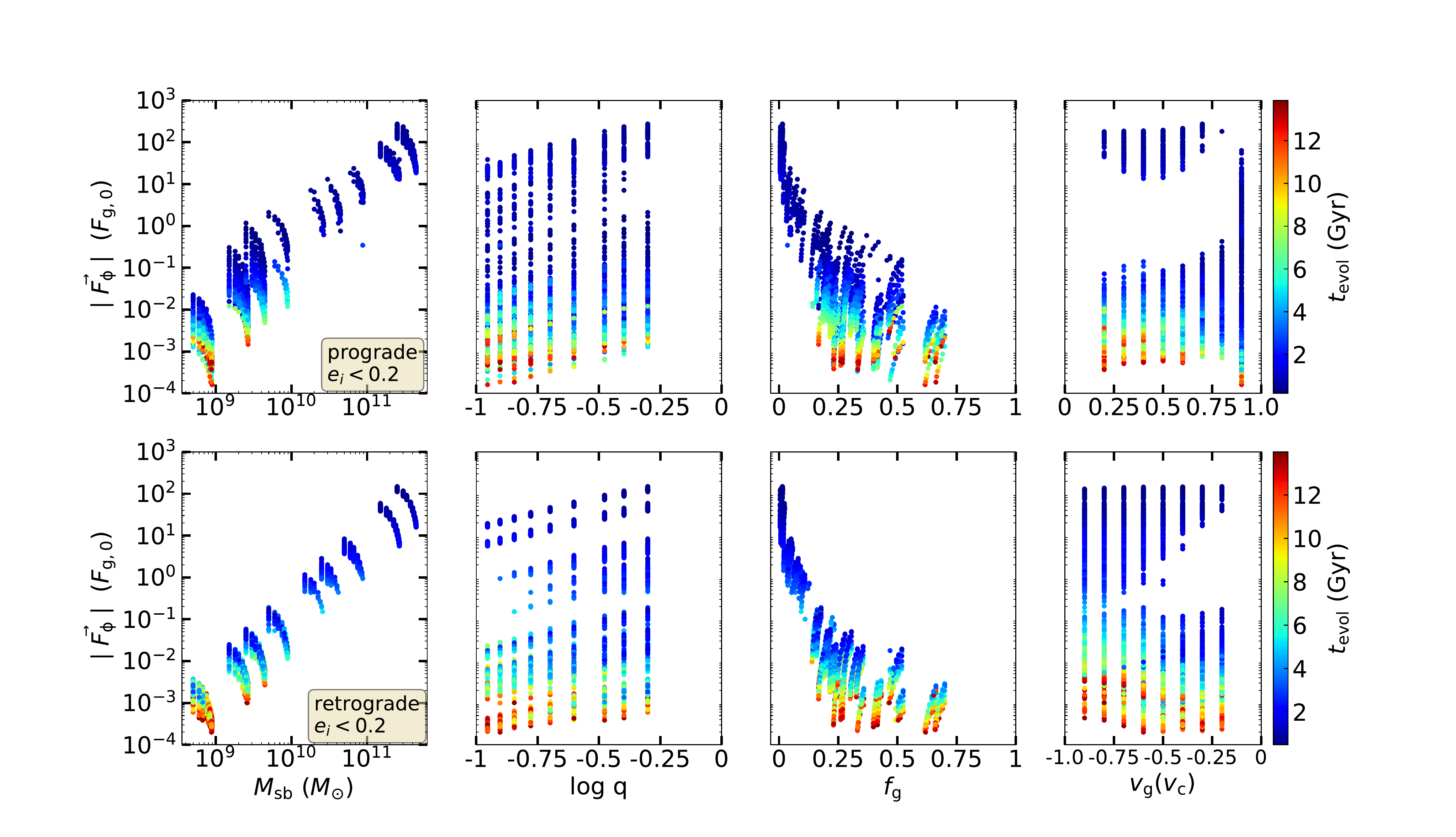}
\caption{The relationship between the three parameters of the model
  ($M_{\rm sb}$, $q$ and \vg), the total DF force, and  the inspiral
  time, \tevol, in models with low initial orbital eccentricity. We
  also introduce the total gas fraction $f_g$ (Eq.~\ref{eq:fg}) to
  more closely connect the model galaxies to observed quantities. We show only the azimuthal component of the DF force which is responsible for the bulk of the orbital evolution of the sMBH. The top (bottom) row of panels correspond to prograde (retrograde) orbits. The color marks the inspiral time.  The deep blue dots with $M_{\mathrm{sb}} > 10^{10}\ {\rm M}_{\odot}$ and $f_{\rm g} \lesssim 0.1$ are bulge dominated cases.}
\label{fig:DFplots-lowe}
\end{figure*}

\begin{figure*}[th]
\centering
     \includegraphics[width=\textwidth]{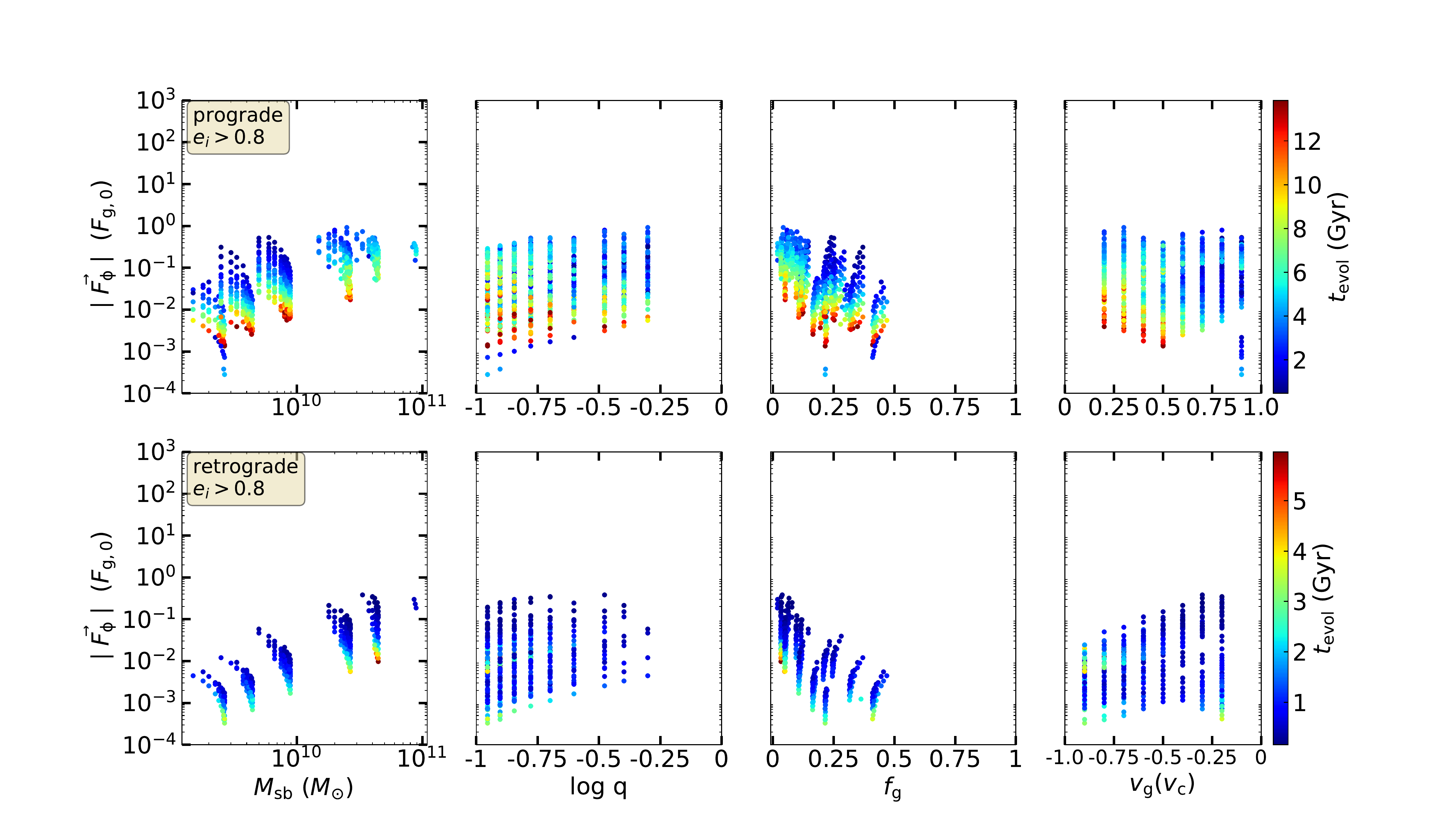}
\caption{Same as Figure~\ref{fig:DFplots-lowe} but now for models with high initial orbital eccentricity. In this group of models, orbital evolution is determined by gas disks and larger $q$ and \vg\ lead to faster inspiral times.}
\label{fig:DFplots-highe}
\end{figure*}

\subsection{The Dependence of Inspiral Time on Host Galaxy and Initial Orbital Properties}
\label{sub:inspiralphysics}

The inspiral time of the sMBH (\tevol) depends sensitively on the total DF force acting on the sMBH and therefore, on the sMBH mass. As a consequence, $q$ has the strongest impact on the inspiral time and we find that $M_{\mathrm{bin}}$, $n_{\rm gd0}$, and $v_{\mathrm{g}}$ also affect it to some degree. We discuss these dependancies in this section. In each simulation we calculate \tevol\ as the time from the start of a simulation to the time when the distance between the pMBH and sMBH is smaller than $1$~pc. 

Figure~\ref{fig:tvsa-lowe} illustrates the dependence of \tevol\ on the mass ratio of the MBH pair. Both panels show the evolution of semimajor axis with time for prograde orbits. The left (right) panel shows the results for $e_{\rm i} < 0.2$ ($e_{\rm i} > 0.8$). In both cases, MBH pairs with larger mass ratios evolve faster than those with smaller ones, in agreement with findings by earlier works \citep{KBH2017, AM2012}. The ankle (left panel) and knee (right) apparent in the curves at $a \ga 0.5$~kpc in Figure~\ref{fig:tvsa-lowe} correspond to the change in the rate of eccentricity evolution shown in Figure~\ref{fig:evsa-lowe} and occur when the Mach number at the apocenter is close to zero. At this point, the sMBH is rotating with the gas disk at nearly the same speed and as a consequence, the influence of the gaseous DF ceases to be important.

Figure~\ref{fig:DFplots-lowe} illustrates how four key parameters of our model ($M_{\rm sb}$, $q$, $f_g$ and \vg) affect \tevol\ for sMBHs with $e_{\rm i} < 0.2$. We first explain the most important aspects of this figure. To quantify the total DF force shown on the $y$ axis, we sum the contributions due to gas disk and stellar disk and bulge for each simulation. We focus only on the dominant, azimuthal component of the force, $\vec{F}_{\phi}$, that is responsible for the orbital evolution and neglect the radial component. The DF force is shown in units of $F_{\rm g,0}= 4\pi\, m_{\rm p}n_{\rm gd0}\,(GM_2/c_{\rm s})^2 = 3.7\times 10^{31}$~dyn, evaluated for $n_{\rm gd0}=100$~cm$^{-3}$, $M_2=10^6$~M$_{\odot}$ and $c_s=10$~km~s$^{-1}$. 

Before considering our ensemble of results, it is useful to introduce the parameter $f_g$ to describe the total gas fraction of the remnant galaxy within the central 1\,kpc
\begin{equation}
  \label{eq:fg}
  f_g=\frac{M_{\mathrm{gas}}}{(M_{\mathrm{gas}}+M_{\mathrm{\star}})}\,,
\end{equation}
where $M_{\mathrm{\star}}$ includes both the mass of the bulge and stellar disk within the central kiloparsec. The value of the gas fraction for a given galaxy model depends on both $n_{\rm gd0}$ and \fgd\ defined before but provides a more intuitive measure of the gas richness of the remnant galaxy. Our main motivation for introducing $f_g$ is to provide the interpretation of our results in terms of the parameter that can be compared directly with the gas fraction of galaxies inferred from observations.

The top and bottom panels in Figure~\ref{fig:DFplots-lowe} represent
configurations characterized by the low eccentricity prograde and
retrograde orbits, respectively. In both cases, the systems in which
the stellar bulge dominates the orbital evolution of the sMBH (i.e.,
when $M_{\rm sb} > 10^{10}\ {\rm M}_{\odot}$ or $f_g \lesssim
0.1$) correspond to the shortest inspiral times, since in these cases
the stellar bulge and gas disk together provide a much stronger DF
force relative to the gas disk alone. Similarly, and as noted before,
the galaxy models with the shortest inspiral time correspond to MBH pairs with higher mass ratios. This is illustrated in the second column of Figure~\ref{fig:DFplots-lowe} for both the prograde and retrograde scenarios, as well as in Figure~\ref{fig:tvsa-lowe}.

\begin{figure*}[t]
  \includegraphics[width=0.49\textwidth]{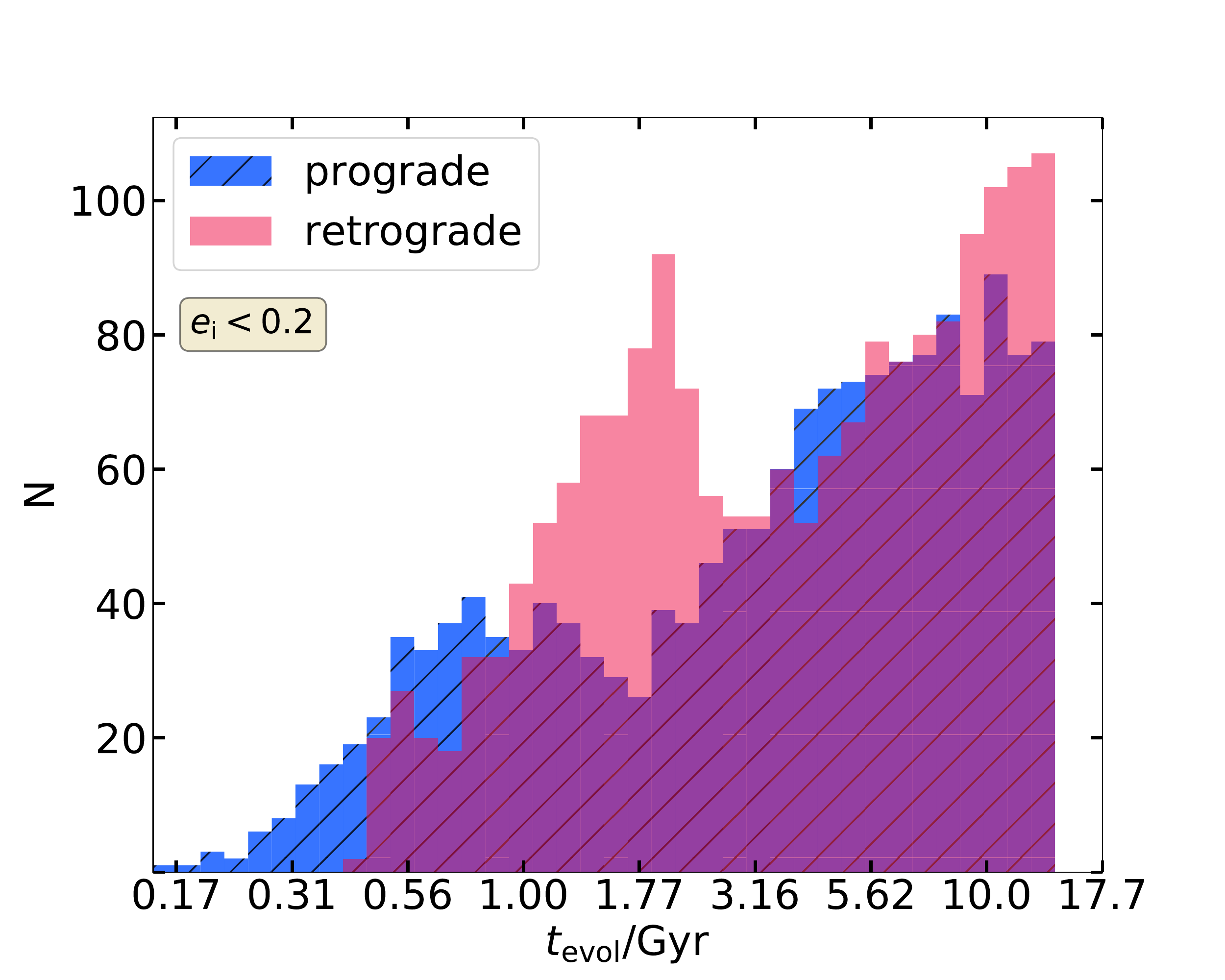}
  \includegraphics[width=0.49\textwidth]{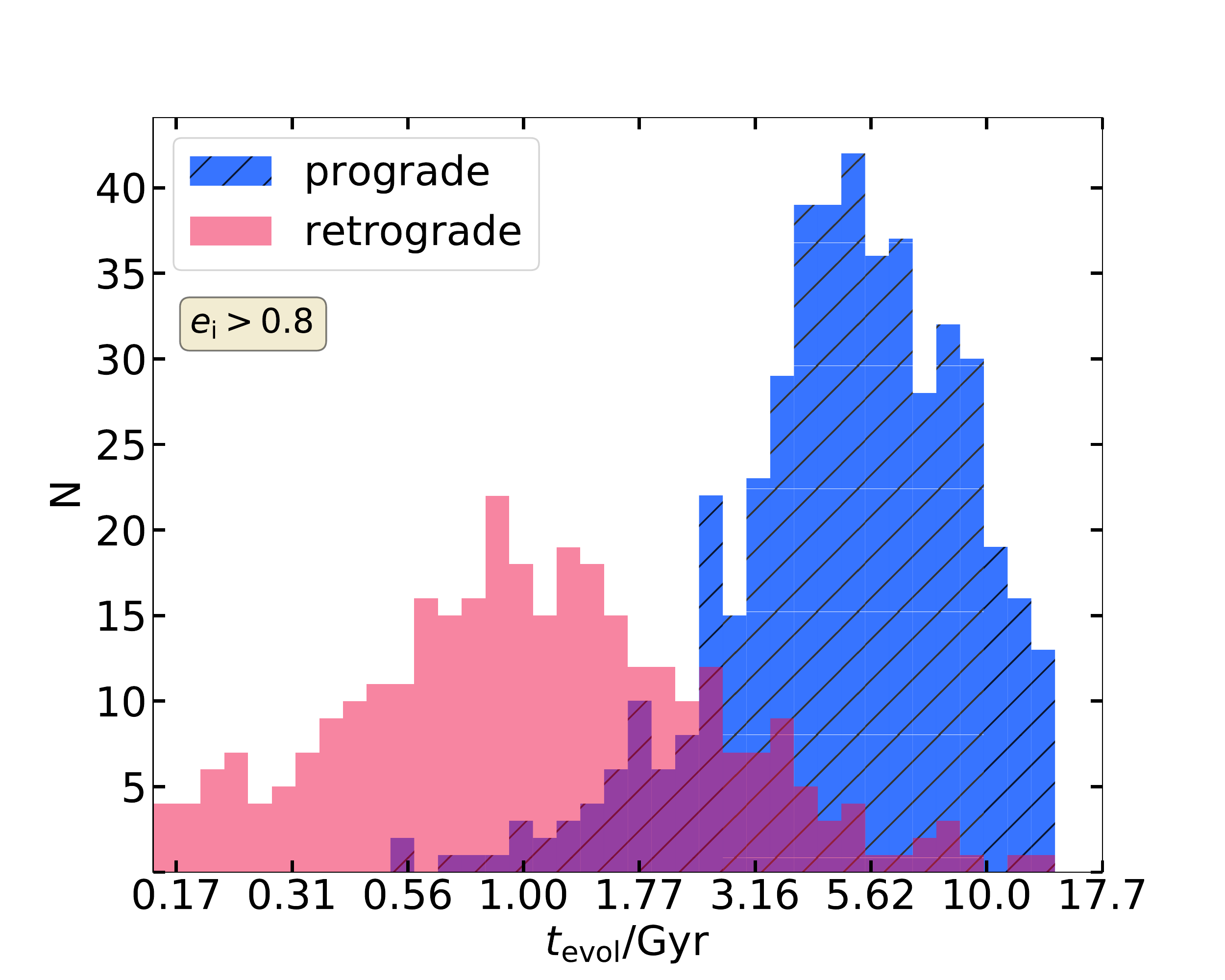}
  \caption{Histograms of \tevol\ derived from the entire model suite for MBHs on prograde and retrograde orbits with low (left) and high $e_{\rm i}$ (right).}
  \label{fig:timehisto-lowe}
\end{figure*}

Alternatively, when the evolution of sMBH is dominated by the gas disk (i.e., $M_{\rm sb} < 10^{10}\ {\rm M}_{\odot}$ or $f_g \gtrsim 0.1$), then the rate of inspiral of a MBH pair depends on the kinematics of the disk. Specifically, the higher values of \vg\ (i.e., close to $v_{\rm c}$) correspond to the shorter inspiral time scales for both prograde and retrograde orbits (see the fourth column in Figure~\ref{fig:DFplots-lowe}). Such gas disks generate a larger gas DF force on sMBHs with low eccentricity orbits, that are themselves orbiting with speeds close to $v_{\rm c}$, which results in a shorter inspiral time. For models in which orbital evolution is predominantly driven by the gas DF, \tevol\ is affected by $M_{\rm sb}$, $q$, and $f_g$ to a lesser degree than in the bulge dominated cases.

Figure~\ref{fig:DFplots-highe} shows how the DF force and \tevol\ are
affected by the same set of galaxy parameters when the sMBH has
$e_{\rm i} > 0.8$. Recall that orbits characterized by high $e_{\rm
  i}$ evolve under the influence of gas disk, because their apocenter
is far outside the stellar bulge. Consequently, \tevol\ is more
strongly affected by \vg\ and less by $M_{\rm sb}$ and $f_g$. The top
row of panels in Figure~\ref{fig:DFplots-highe} shows that when the
sMBH is in a prograde orbit, galaxies with larger \vg\ and $q$ lead to
a faster inspiral. For retrograde orbits with high $e_{\rm i}$,
neither the bulge nor the gas disk provide enough DF force to
circularize the orbit. Thus, the eccentricity remains very high ($\sim
0.9$), which makes the orbital pericenter pass below 1 pc (our
stopping criterion) much sooner than in other simulations. Indeed,
when $e_{\rm i}$ is high, it is possible for the sMBH to plunge toward the primary MBH if the eccentricity continues to increase.

Figure~\ref{fig:timehisto-lowe} shows histograms of \tevol\ for the
entire model suite found for different types of MBH orbits, starting with low and high initial eccentricity. Orbits with $e_{\rm i} < 0.2$ have a bimodal distribution of \tevol\ (left panel of the same figure). The first peak of the histogram of the prograde orbits (at $\sim 1$\,Gyr) corresponds to models in which the stellar bulge dominates the orbital evolution  of the sMBH. The second peak (at $\sim 5$\,Gyr) of the same histogram corresponds to gas disk dominated orbital evolution.  The retrograde orbits with $e_{\rm i} < 0.2$ (also shown in the left of Figure~\ref{fig:timehisto-lowe}) have the first peak at $\sim 1.77$\,Gyr and second peak beyond $\sim 10$\,Gyr. In this case too, models in which orbital evolution is predominantly driven by the bulge DF have shorter inspiral times than those whose evolution is determined by the gas disk, due to a significantly larger DF force exerted by the bulge (see Figure~\ref{fig:DFplots-lowe}).

The right panel of Figure~\ref{fig:timehisto-lowe} shows histograms of \tevol\ for sMBH orbits with high initial eccentricity. In this group of models, a large fraction of prograde orbits have an inspiral time close to 5\,Gyr years. The eccentric orbits take the MBHs outside of the radius of influence of the stellar bulge, leaving the gas disk as the main contributor to the DF force at the orbital apocenter. As a result, the inspiral time of prograde orbits is long. The eccentricity of the retrograde orbits on the other hand continues to increase, until the sMBH plunges into the 1\,pc radius, after about 1\,Gyr. Note again, not all of these eccentric retrograde orbits will lead the sMBH to plunge into the pMBH. Some sMBHs in eccentric retrograde orbits may reverse their direction and enter the circularization phase as mentioned in Sec~\ref{S_pro_ret} before they merge with the pMBHs. We will be able to give a more precise prediction in the future, when the resolution of our simulation is improved.

\section{Discussion}
\label{sec:discuss}

\begin{figure*}[t]
\centering
        \begin{tabular}{@{}cc@{}}
            \includegraphics[width=0.49\textwidth]{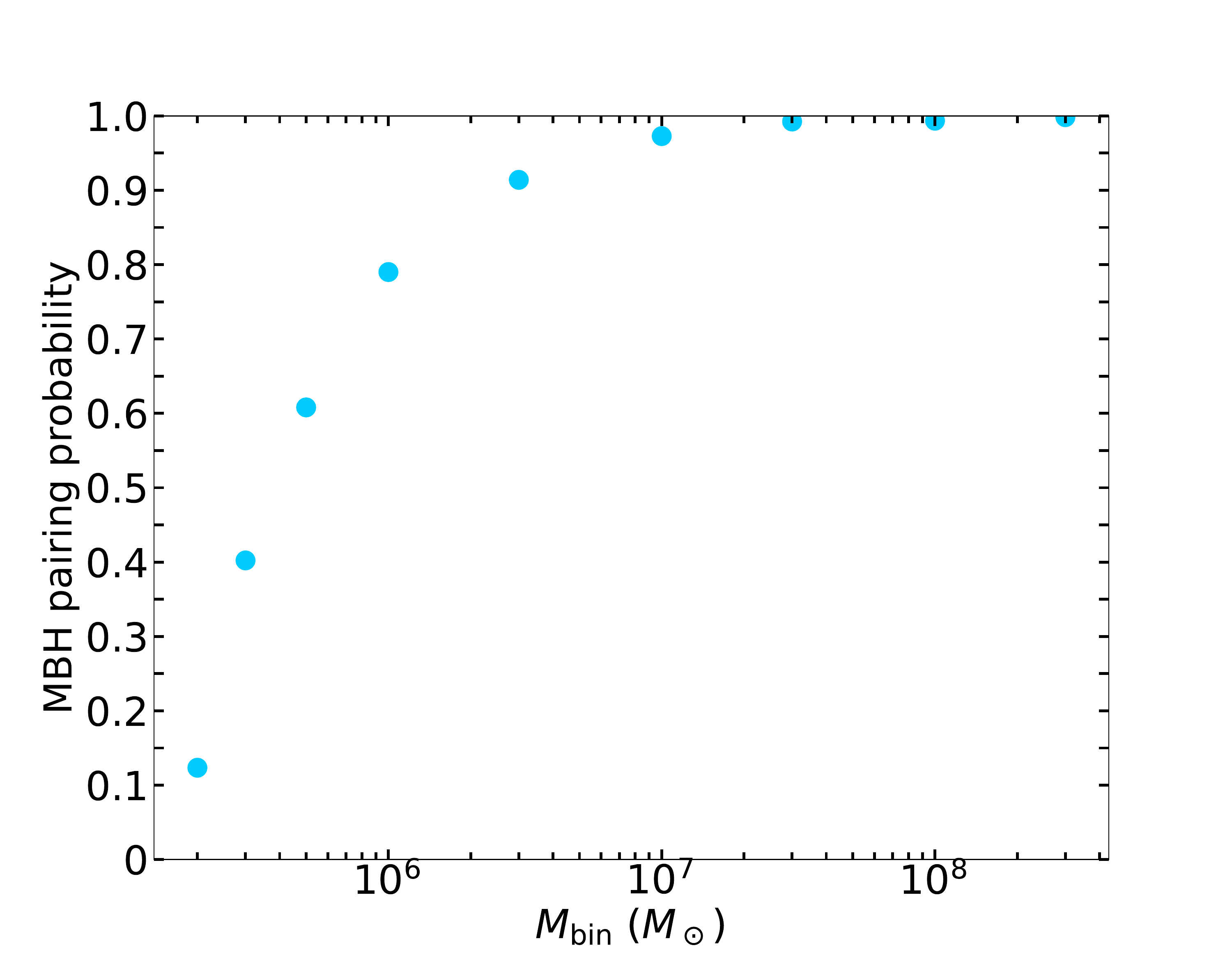}
            \includegraphics[width=0.49\textwidth]{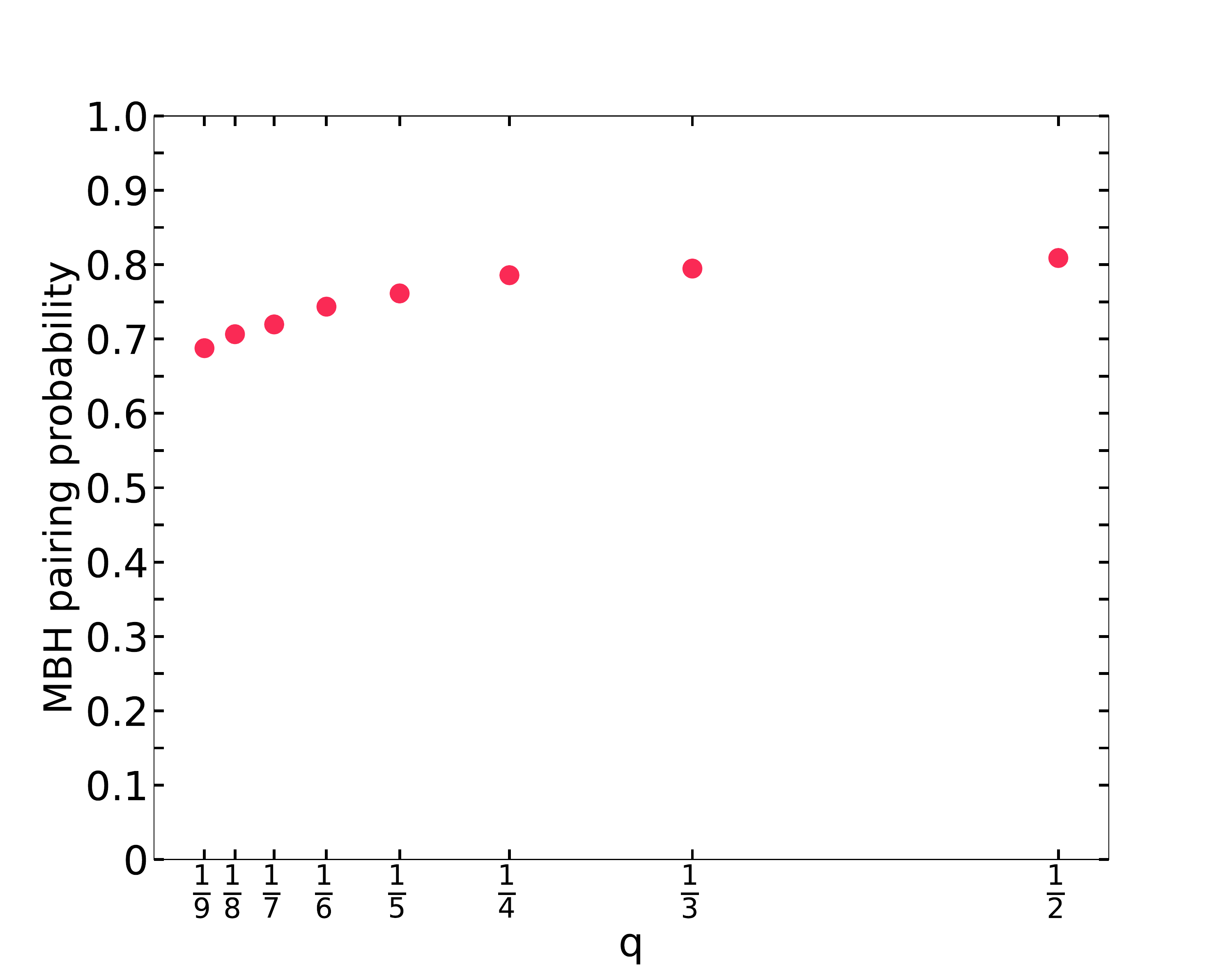}\\
            \includegraphics[width=0.49\textwidth]{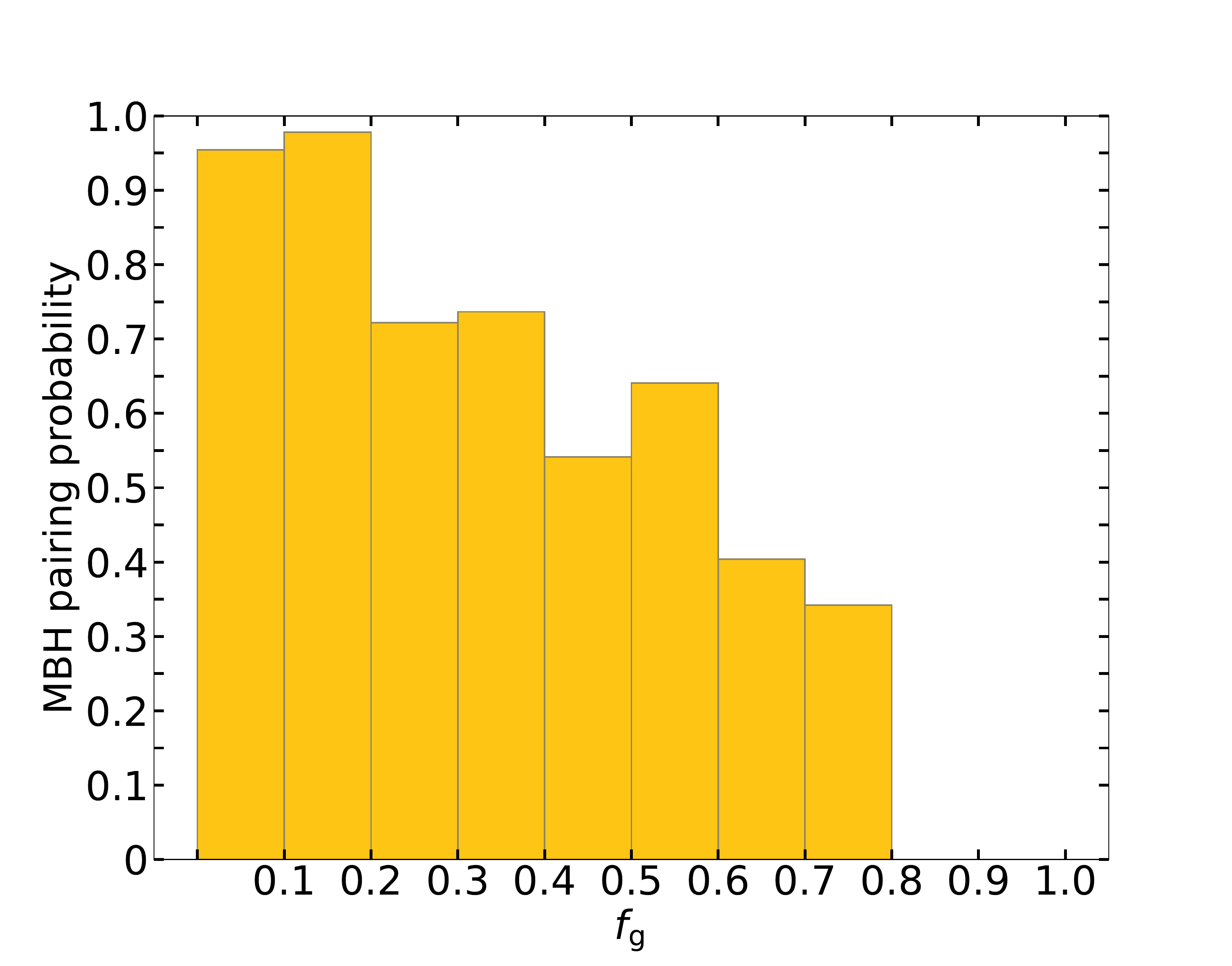}
            \includegraphics[width=0.49\textwidth]{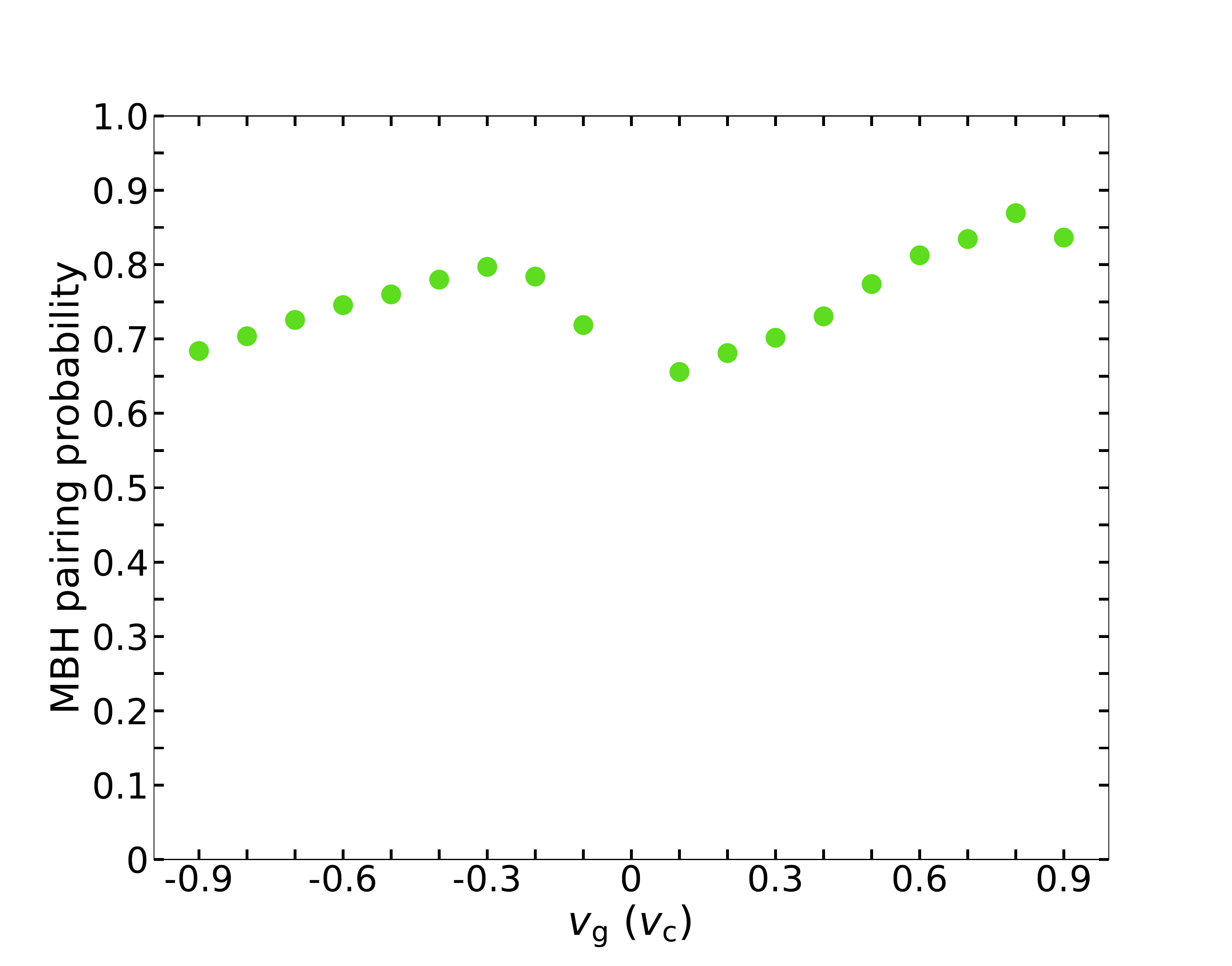}
        \end{tabular}
\caption{MBH pairing probability as a function of the host galaxy and MBH pair properties: \Mtot (top left), q (top right), $f_{\rm g}$ (bottom left), and \vg\ (bottom right). We show the dependence on $f_{\rm g}$ as a histogram, since this is a derived, rather than a primary parameter with assumed equidistant values.}
\label{fig:merger_rate}
\end{figure*}

The results described above show how the eccentricity of the orbit and the inspiral time are influenced by the properties of the host galaxy and the MBH pair itself. In this section, we describe the implications of these results for understanding the evolution of the MBH pairs and for searches for dual AGNs. We also discuss the impact of simplifying assumptions used in our model.

\subsection{The Probability of MBH Pairing in Different Remnant Galaxies}
\label{sub:galaxyprop}

To investigate which remnant galaxies are more likely to form gravitationally bound MBH binaries that may evolve to coalescence, we calculate the probability of MBH pairing as a function of \Mtot, $q$, $f_{\rm g}$, and \vg. The MBH pairing probability is defined as the fraction of MBH pairs that have $t_{\rm evol}$ shorter than a Hubble time. We assume that a MBHB is likely to form once the MBH pair finds itself at a separation smaller than 1 pc. The pairing probability calculated in this way is an upper limit, because not all eccentric binaries that reach separations below 1 pc are gravitationally bound. The top left panel of Figure~\ref{fig:merger_rate} illustrates that MBH pairing probability increases when $M_{\rm bin}$ increases, and the pairing probability becomes 1 when $M_{\rm bin} \geqslant 3\times 10^7 M_{\rm \odot}$. In comparison, MBH pairs in galaxies with $M_{\rm bin} \leqslant 5\times10^5 M_{\rm \odot}$ have an average MBH pairing probability of $\sim 0.5$. The MBH pairing probability increases steeply from $\sim 0.1$ to nearly 1.0 in the range of $2\times 10^5 \lesssim M_{\rm bin} \lesssim 10^7 M_{\rm \odot}$.

The top right panel of the same figure shows the MBH pairing probability as a function of the mass ratio. The MBH pairing probability grows with the mass ratio and reaches a maximum value of 0.8 at $q = 1/2$. The true pairing probability is likely underestimated by our model, in which the primary MBH is fixed to the center of the galaxy. The assumption of a fixed pMBH is more appropriate for unequal mass pairs with $q< 1/4$ and thus, we expect that the pairing probability for these values of the mass ratio is more realistic. The variation in the MBH pairing probability with $q$ is less than 20\% for the full range of mass ratios explored in this calculation, indicating that its impact is weaker than that of the other galaxy properties. 

The bottom left panel of  Figure~\ref{fig:merger_rate} shows that the rate at which MBH pairs form in galaxies with $f_{\rm g} < 0.2$ is $\sim 1$, two times larger than the pairing probability of galaxies with $f_{\rm g} > 0.6$. This indicates that galaxies with dominant stellar bulges are more likely to host GW sources than the gas dominated ones. The bottom right panel illustrates how the MBH pairing probability depends on \vg. The pairing probability peaks at $v_{\rm g}=0.8 v_{\rm c}$ and $v_{\rm g}=-0.3 v_{\rm c}$. The peak at $v_{\rm g}=0.8 v_{\rm c}$ is due to MBH pairs in circular orbits that experience efficient gaseous DF. The peak at $v_{\rm g}=-0.3 v_{\rm c}$ is due to MBH pairs in eccentric orbits whose eccentricity continues to increase, which results in the sMBH plunging into the central parsec. It is worth noting that the pairing probability of MBHs depends sensitively on the magnitude of \vg. This dependence on the kinematic properties of the gas disk, where efficient pairing is a possibility but not a guaranteed outcome, is responsible for the suppression in the average MBH pairing rate in the bottom left panel, evident for higher values of $f_{\rm g}$, when a gas disk dominates the orbital evolution of the pair.

In summary, the merger remnant galaxies with the MBH pairing probability larger than 80\% have at least one of the following properties: (a) MBH pair with mass larger than $10^6 M_{\rm  \odot}$, (b) MBH mass ratio larger than $1/4$, (c) gas fraction smaller than  $0.2$, (d) a gas disk corotating with the sMBH with $v_{\rm g} \geq 0.5 v_{\rm c}$, and (e) a gas disk counterrotating with the sMBH with $v_{\rm g} = - 0.3 v_{\rm c}$.  It is also worth noting that the initial eccentricity does not affect the MBH pairing probability significantly. Models characterized by orbits with low $e_{\rm i}$ have a MBH pairing probability of $0.62$, nearly indistinguishable from  that for models with high $e_{\rm i}$ orbits, which is $0.67$.

On the other hand, galaxies more likely to host dual AGNs are the ones where the inspiral time is long and the two MBHs spend more time at large separations. Thus, dual AGNs are more likely to be found in
galaxies with at least one of the these characteristics: (a) a pair mass smaller than $5\times 10^5 M_{\rm \odot}$, (b) mass ratio smaller than $1/4$, (c) a gas fraction larger than $0.2$, (d) a gas disk corotating with the sMBH with $v_{\rm g} \leq 0.5 v_{\rm c}$, and (e) a gas disk counterrotating with the sMBH with $v_{\rm g} \ne 0.3 v_{\rm c}$. All other things being the same, MBH pairs on orbits with high eccentricity are also more likely to be observed as dual AGNs relative to the low eccentricity systems, as the two MBHs spend a lot of time far apart and close to the orbital apocenter.
 
Our findings are in good agreement with those of other groups, including both hydrodynamical simulations and semi-analytic models \citep{L2008, C2015, T2016, K2019}. Specifically, these works find that the inspiral time of low initial eccentricity MBHBs is shorter in host galaxies with larger bulge mass, larger MBH mass ratio, and either gas or stellar disk rotating along with the secondary MBH with velocity close to $v_{\rm c}$. In a recent work, \citet{K2019} perform 20 N-body simulations to study the evolution of MBHBs under the influence of DF and three-body scattering in a stellar disk. Although they do not account for the gas DF, they find that prograde orbits decay faster than the retrograde, which is also in agreement with our results.


\subsection{Impact of Simplifying Assumptions}
\label{sub:assumptions}

The power of using the semi-analytic approach is the ability to compute a large number of simulations of MBH orbital decay, over a wide range of galaxy and MBH properties but at the cost of making some simplifying
assumptions. We describe the impact of these assumptions in this section.

In this work, we assume that the pMBH is fixed at the center of the host galaxy, even for MBH pairs with a mass ratio of $q=1/2$. If the motion of pMBH and its associated DF force were modeled in the simulations, the resulting inspiral times for such pairs would be shorter. Consequently, we expect that we overestimate the inspiral time in MBH pairs with comparable masses. Similarly, we make a simplifying assumption that the two MBHs do not grow through accretion over the length of the simulation. Have they been able to do so, the increase in the total mass of the binary would render the inspiral time shorter.


The models calculated in this work also assume a ``bare" sMBH, completely stripped of the nuclear star cluster. This is a plausible outcome and \citet{KBH2017} show that by the time a MBH pair reaches a separation of 1~kpc (the starting point of our simulations), there is a high probability that the sMBH has been stripped of any bound stellar cluster. If nevertheless there is a surviving remnant of the stellar cluster around the sMBH, this would result in more efficient DF and shorter inspiral time scales. 



We make an assumption that the sMBH is on a co-planar orbit with the gas and stellar disks in all our models. In the case that the sMBH was on an inclined orbit relative to the gas and stellar disks of the galaxy, the DF exerted by the stellar bulge would remain the leading mechanism for orbital evolution. The inspiral timescale would then depend on the exact properties of the bulge, as well as the properties of the MBH orbit. \citet{VW2014} highlight an additional effect for MBHs on inclined orbits: inclined orbits take the secondary MBH outside of the gas disk and in such way avoid rapid stripping all of its own gas reservoir due to the ram pressure. In this scenario, the pericentric passage of the sMBH through the galaxy remnant can trigger the star formation, resulting in a formation of a dense stellar cusp around the sMBH. This effect was found to increase the mass of the secondary nucleus and as a result, shorten the inspiral time \citep{VW2014}. 

An additional feature of our model is an assumption of a smooth gas disk without spiral arms, gas clumps and inhomogeneities. In reality, the interaction between the secondary MBH and giant molecular clouds or stellar clusters can cause a random walk of the orbit of the sMBH, leading to a slow inspiral or stalling. The interaction with strong spiral density waves can even eject the sMBH from the disk plane or slow down the decay by orders of magnitude \citep[e.g.,][]{T2016}.


Although several of the effects mentioned above may decrease \tevol, radiation feedback arising from the accreting MBH has been shown to render the gas DF inefficient for some range of scenarios \citep{PB2017}. 
Intriguingly, the radiation feedback may even change the direction of the DF force, speeding up the sMBH \citep{Err2019}. If so, we expect the inspiral time to increase in the presence of radiation feedback for some of our model configurations. Adding the effect of radiation feedback to these simulations will be the subject of a future paper. 

\section{Conclusions}
\label{sec:concl}

We present the results of a semi-analytic model for orbital evolution of a MBH pair in a merger remnant galaxy, under the influence of stellar and gas dynamical friction. The model describes the evolution of unequal mass pairs from initial separations of about a kiloparsec to $<1$ pc, where they are likely to form gravitationally bound MBH binaries. We use a grid of nearly 40,000 configurations to investigate how the pairing of MBHs is affected by the properties of the host galaxy and the two MBHs. Our main findings are as follows:

\begin{itemize}
\item Orbital eccentricity is the key parameter that determines the efficiency of DF and consequently, a pairing probability of MBHs in galaxies. We find that the evolution of initially low eccentricity orbits can be dominated by gas disks (when gas fraction in the galaxy is larger than $0.1$) or stellar bulges (when gas fraction is smaller than $0.1$). Evolution of initially highly eccentric orbits is always dominated by gas disks, since the secondary MBH spends most of its time outside of the region where it can be affected by the bulge. We find the contribution to the DF force from the stellar disk always to be smaller than, and negligible with respect to those of the bulge and the gas disk in our model galaxies. 
  
\item Orbits of MBHs evolving under the influence of the stellar bulge, which does not have a coherent rotational motion, circularize and shrink in size. When the DF force is dominated by the gas disk however, the orbital evolution of the secondary MBH sensitively depends on the relative speed between the gas disk and the sMBH, and does not lead to circularization in general. Specifically, we find that the final eccentricity of the secondary MBHs that corotate with the the gas disk is inversely proportional to the gas rotational speed. The secondary MBHs on retrograde orbits however can be driven to high eccentricities by the gas disk. 
  
\item The chance of MBH pairing within a Hubble time is higher than 80\% in host galaxies with a gas fraction $< 20$\%. This is because efficient pairing is a possibility but not a guaranteed outcome in scenarios when the gas disk dominates the orbital evolution of the MBHs, resulting in a lower average incidence of close MBH pairs compared to the scenarios in which the evolution is influenced by the stellar bulge. The pairing probability is equally high in galaxies hosting MBH pairs with total mass $ > 10^6 M_{\rm \odot}$ and mass ratios $\geq1/4$. 

\item The formation of close separation MBH pairs tends to be fastest in galaxies with one or more of the following properties: large stellar bulge, comparable mass MBHs, and the galactic scale gas disk with rotational speed close to circular. In these galaxies, MBHs on circular prograde orbits and eccentric retrograde orbits have the shortest inspiral times and are therefore  the most likely progenitors of MBH binaries that coalesce due to the emission of GWs. Conversely, the galaxies with the opposite properties, that host slowly evolving MBH pairs, are the most likely hosts to dual AGNs observed at kiloparsec separations.
  
\end{itemize}


Presently, the observations of dual AGNs are uncovering more and more
late-stage galaxy mergers, paving the way toward the regime where
dynamical friction is expected to shape the evolution of the MBH
pairs. Based on the progress so far, we anticipate that in the not too
distant future we will be able to test the nature and efficiency of
dynamical friction by directly comparing  the observations to models like the
one presented in this work. More generally, such comparisons will
reveal which galaxies are more likely to host MBH pairs at a
particular redshift, and how these properties evolve with time. This
knowledge will be crucial for the future and present electromagnetic
and GW observatories, for it will indicate where to look for possible MBH mergers. 
 
\acknowledgments

T.B. acknowledges the support by the National Aeronautics and Space Administration (NASA) under award No. 80NSSC19K0319 and by the National Science Foundation (NSF) under award No. 1908042. 

\bibliography{refs}

\bibliographystyle{aasjournal}

\end{document}